\documentclass[fleqn,usenatbib]{mnras}
\usepackage[T1]{fontenc}
\usepackage{ae,aecompl}
\usepackage{graphicx}
\usepackage{amsmath}
\usepackage{amssymb}
\usepackage{mathrsfs}
\usepackage{booktabs}
\usepackage{threeparttable}

\pubyear{2021}

\begin{document}

\title[The Photosphere Spectrum of Hybrid Jet]{The Photosphere Emission
Spectrum of Hybrid Relativistic Outflow for Gamma-ray Bursts}
\author[Meng et al.]{ Yan-Zhi Meng,$^{1,2,3}$\thanks{%
E-mail: yzmeng@nju.edu.cn (YZM)} Jin-Jun Geng$^{1}$\thanks{%
E-mail: jjgeng@pmo.ac.cn (JJG)} and Xue-Feng Wu$^{1}$\thanks{%
E-mail: xfwu@pmo.ac.cn (XFW)} \\
$^{1}$Purple Mountain Observatory, Chinese Academy of Sciences, Nanjing
210023, China \\
$^{2}$School of Astronomy and Space Science, Nanjing University, Nanjing
210023, China\\
$^{3}$Key Laboratory of Modern Astronomy and Astrophysics (Nanjing
University), Ministry of Education, China }
\date{Accepted XXX. Received YYY; in original form ZZZ}
\maketitle

\begin{abstract}
The photospheric emission in the prompt phase is the natural prediction of
the original fireball model for gamma-ray burst (GRB) due to the large
optical depth ($\tau >1$) at the base of the outflow, which is supported by
the quasi-thermal components detected in several \textit{Fermi} GRBs.
However, which radiation mechanism (photosphere or synchrotron) dominates in
most GRB spectra is still under hot debate. The shape of the observed
photosphere spectrum from a pure hot fireball or a pure
Poynting-flux-dominated outflow has been investigated before. In this work,
we further study the photosphere spectrum from a hybrid outflow containing
both a thermal component and a magnetic component with moderate
magnetization ($\sigma _{0}=L_{P}/L_{\text{Th}}\sim 1-10$), by invoking the
probability photosphere model. The high-energy spectrum from such a hybrid
outflow is a power law rather than an exponential cutoff, which is
compatible with the observed Band function in a great amount of GRBs. Also,
the distribution of the low-energy indices (corresponding to the peak-flux
spectra) is found to be quite consistent with the statistical result for the
peak-flux spectra of GRBs best-fitted by the Band function, with similar
angular profiles of structured jet in our previous works. Finally, the
observed distribution of the high-energy indices can be well understood
after considering the different magnetic acceleration (due to magnetic
reconnection and kink instability) and the angular profiles of dimensionless
entropy with the narrower core.
\end{abstract}

\label{firstpage} \pagerange{\pageref{firstpage}--\pageref{lastpage}}


\section{INTRODUCTION}

After decades of researches, the radiation mechanism of GRB prompt emission
is still unclear %
\citep[e.g.,][]{ZhaYan2011,ZhaBo2014,Geng18,Geng2019,Lin2018,ZhaBB2018,ZhangBB18b,ZhangBB2021,Duan2019,Duan2019b,Huang2019,Li2019b,Yang2020,ZhaB2020}%
. The photospheric emission model seems to be a promising scenario %
\citep[e.g.,][]{Abra1991,Thom1994,Me2000,Ree2005,Naga2011,Pe2011,Fan2012,Lazz2013,Ru2013,Gao2015,Be2015,Pe2015,Ry2017,Acun2018,Hou2018,Meng2018,Meng2019,Li2019a,Li2019c,Li2019d,Acun2020,Wang2020,Wang2021}%
. The photospheric emission is the prediction of the original fireball model %
\citep{Good1986,Pac1986}, because the optical depth $\tau $ at the outflow
base is much greater than unity \citep[e.g.][]{Pi1999}. As the fireball
expands and the optical depth decreases, the internally trapped photons
finally escape at the photosphere radius ($\tau =1$). The photospheric
emission model naturally interprets the clustering of the peak energies %
\citep[e.g.,][]{Pree2000,Kan2006,Gold2012} and the high radiation efficiency %
\citep{Lloy2004,ZhaB2007,Wygo2016} observed.

Indeed, based on the analyses of the observed spectral shape, a
quasi-thermal component has been found in a great amount of BATSE GRBs %
\citep{Ry2004,Ry2005,Ry2009} and several \textit{Fermi} GRBs (GRB 090902B, %
\citealt{Abdo2009}, \citealt{Ry2010}, \citealt{ZhaB2011}; GRB100724B, %
\citealt{Gui2011}; GRB 110721A, \citealt{Axel2012}; GRB 100507, %
\citealt{Ghir2013}; GRB 101219B, \citealt{Lar2015}; and the short GRB
120323A, \citealt{Gui2013}). Moreover, in GRB 090902B, the quasi-thermal
emission dominates the observed emission. However, whether the whole
observed Band function or cutoff power law (the COMP model) can be explained
by the photosphere emission alone remains unknown. Some statistic aspects of
the spectral analysis results for large GRB sample seem to support this
point. First, some observed bursts have a harder low-energy spectral index
than the death line $\alpha $= $-$2/3 of the basic synchrotron model 
\footnote{The so-called synchrotron line-of-death can be violated
both from the data analysis point of view\ \citep{Burgess2020} and the
theoretical point of view \citep{Yang2018}.}, especially for the peak-flux
spectrum and short GRBs %
\citep[e.g.,][]{Kan2006,ZhaBB2011,Gold2012,Gold2013,Grub2014,Yu2016,Bur2017,Lu2017}%
. Second, the\ cutoff power law is the best-fit spectral model for more than
a half of the GRBs, which is a natural expectation within the photosphere
emission model. At last, for a large fraction of GRBs, the spectral width is
found quite narrow\ \citep{AxBo2015,Yu2015}\footnote{It was found by
independent groups \ \citep{Zhang2016,Burgess2019} that the spectra are not
necessarily narrow. The same data set could be used to match different input
spectral shapes and the reason for the narrow spectrum claim was that the
Band function itself is narrow. The same data set could be equally fitted
with the broader synchrotron spectra.}. To account for the broad spectra of
other GRBs, quasi-thermal spectrum need to be broadened. Two different
mechanisms of broadening have been proposed theoretically, i.e., the
subphotospheric dissipation \citep{Ree2005,Gian2006,Vur2016,Belo2016} and
the geometric broadening \citep{Pe2008,Lund2013,Deng2014,Meng2018,Meng2019}.

Pure photosphere is considered as that all the photons perform the last
scattering at the radius where the Thompson scattering optical depth drops
down to unity ($\tau =1$). The geometric broadening, namely the probability
photosphere, is the result of the fact that photons can be last scattered at
any place $(r,\Omega)$ inside the outflow if only the electron exists there,
where $r$ is the distance away from the explosion center and $\Omega (\theta
,\phi)$ is the angular coordinate. Therefore, a probability density function 
$P(r,\Omega)$ is introduced to describe the probability of last scattering
at any position \citep{Pe2008,Pe2011,Belo2011}. Then, the observed
photosphere spectrum is the superposition of a series of blackbodies with
different temperature, thus broadened.

Considering the geometric broadening, \citet{Deng2014} studied detailed
photosphere spectrum for a spherically symmetric wind. They found that the
low-energy spectrum can be modified to $F_{\nu }\sim \nu ^{1.5}$ ($\alpha
\sim +0.5$), but still much harder than the typical observation ($\alpha
\sim -1.0$). On the other hand, the anticorrelation between the peak energy $%
E_{p}$ and the photosphere luminosity $L_{\text{ph}}$ seems to be in
conflict with the observed hard-to-soft evolution or the intensity tracking
for $E_{p}$~\citep{Lia1996,Ford1995,Ghir2010,Lu2010,Lu2012}. Thus, a more
complicated photosphere model is needed. By concerning on the photosphere
emission from a jet with a specific angular structure, the observed typical
low-energy photon index $\alpha \sim -1.0$ (see also \citealt{Lund2013}) and 
$E_{p}$ evolutions could both be reproduced~\citep{Meng2019}. For long GRBs
from collapsars~\citep{Mac1999}, the jet is collimated by the pressure of
the surrounding gas when propagating through the collapsing progenitor star~%
\citep[e.g.][]{ZhaWoo2003,Mor2007,Mizu2011}, hence it could have angular
profiles of energy and Lorentz factor, namely a structured jet %
\citep[e.g.,][]{Dai2001,Rossi2002,ZhMe2002,Kumar2003}. For short GRBs, the
structured jet has also been favored by many theoretical %
\citep[e.g.,][]{Sap2014} and numerical %
\citep[e.g.,][]{Aloy2005,Tche2008,Komi2010,Ross2013,Murgu2017,Gott2020}
studies. Besides, the resulted distribution of the low-energy spectral
indices and the spectral evolution for GRBs best-fitted by the cutoff
power-law model in this structured jet scenario could be consistent with
observed ones.

In studies on the probability photosphere model till now, the jet is
accelerated solely by the radiative pressure of the thermal photons, which
means the jet is dominated by the thermal energy. However, several works on
the central engine reveal that the jet often consists of two components: a
thermal component from the neutrino heating of the accretion disk around the
black hole or the proto neutron star, and a magnetic component (Poynting
flux) launched from the magnetosphere of the central engine %
\citep[e.g.][]{Met2011,Lei2013}. Thus, the magnetically driven acceleration %
\citep{Dren2002,DrenSpru2002,Lyu2003,Gian2006,GianSpru2006,Me2011} may also
play an important role in the real situation, which has not been properly
considered in the framework of the probability photosphere model. In this
work we study the photospheric emission spectrum within the framework of the
probability photosphere model for a hybrid relativistic outflow, which
contains a thermal component and a magnetic component, including both the
thermally and magnetically driven acceleration.

Note that there are significant differences between our work and the study
of photosphere emission from a hybrid relativistic outflow by \cite{Gao2015}%
. On one hand, \cite{Gao2015} only considered a pure blackbody for the
thermal component \citep{Ree1994,Me2002}, rather than the probability
photosphere here. On the other hand, the main purpose of their work is to
compare the flux of the blackbody and that of the non-thermal component, for
hybrid outflows with different compositions. While our work focuses on the
shape of the observed spectra for different hybrid outflow.

The paper is organized as follows. In Section \ref{sec:calcu}, we describe
the calculations of the photospheric emission spectrum for a hybrid outflow
within the probability photosphere model, including the energy injections of
impulsive injection and more reasonable continuous wind. The calculated
spectral results and parameter dependencies are shown in Section \ref%
{sec:result}. In Section \ref{sec:dis}, we discuss the influence of magnetic
dissipation and the radiative efficiency for our photosphere model, %
the assumption of angle-independent luminosity, and impact of the
synchrotron emission. The conclusions are summarized in Section \ref%
{sec:conclu}.

\section{PROBABILITY PHOTOSPHERE EMISSION FROM A HYBRID JET}

\label{sec:calcu}

\subsection{Hybrid Jet and Its Dynamics}

A hybrid jet is composed of a thermal component and a magnetic component,
which could be described by the dimensionless entropy $\eta$ and the initial
magnetization parameter $\sigma _{0}$. The dimensionless entropy $\eta$
represents the average energy per baryon (including the rest mass energy and
the thermal energy) for the thermal component, and the magnetization
parameter $\sigma _{0}$ is the ratio of the magnetic component to the
thermal component, i.e., 
\begin{equation}
\sigma _{0}\equiv \frac{L_{P}}{L_{\text{Th}}}=\frac{L_{P}}{\eta \dot{M}c^{2}}%
,
\end{equation}
where $L_{\text{Th}}$ and $L_{P}$ are the luminosities of the thermal
component and the magnetic (Poynting flux) component, respectively.

Within a hybrid jet, both the radiative pressure and the magnetic pressure
gradient are responsible for the acceleration of the jet. The thermally
driven acceleration proceeds very rapidly with a linear acceleration law $%
\Gamma \propto $ $r$, where $\Gamma$ is the bulk Lorentz factor. And the magnetically driven acceleration proceeds also
rapidly with an acceleration law close to linear $\Gamma \propto $ $%
r^{\lambda }$ ($\lambda =1/2\sim 1$, \citealt{Komi2009,Gran2011}) below the
magneto-sonic point, where the bulk Lorentz factor equals the
\textquotedblleft Alfv\'{e}nic\textquotedblright\ Lorentz factor $\Gamma
_{A}=(1+\sigma )^{1/2}$. Above the magneto-sonic point, the acceleration
proceeds relatively more slowly approximately as $\Gamma \propto $ $%
r^{\delta }$ \citep{Dren2002,DrenSpru2002,Me2011,Vere2012}. Thus, similar to %
\citet{Gao2015}, we approximately assume that the jet is accelerated
linearly until the rapid acceleration radius $R_{\text{ra}}$, which is the
larger one of the thermal saturated radius and the magneto-sonic point, and
then undergoes a slower acceleration with $\Gamma \propto $ $r^{\delta }$
between $R_{\text{ra}}$ and the coasting radius $R_{c}$. Namely, the
dynamics for a hybrid jet may be approximated as (see Figure 1): 
\begin{equation}
\Gamma (r)=\left\{ 
\begin{array}{l}
\frac{r}{r_{0}},\text{ \ \ \ \ \ \ \ \ \ \ \ \ \ \ \ \ \ \ }r_{0}<r<R_{\text{%
ra}}; \\ 
\Gamma _{\text{ra}}\text{ }(\frac{r}{R_{\text{ra}}})^{\mathbf{\delta }},%
\text{\ \ \ \ \ \ \ \ }R_{\text{ra}}<r<R_{c}; \\ 
\Gamma _{c},\text{ \ \ \ \ \ \ \ \ \ \ \ \ \ \ \ \ \ \ }r>R_{c},%
\end{array}%
\right.  \label{b}
\end{equation}%
where $r_{0}$ is the radius at the jet base, $\Gamma _{\text{ra}}=\max (\eta
,$ $[\eta (1+\sigma _{0})]^{1/3})$ is the Lorentz factor at $R_{\text{ra}}$, 
$\Gamma _{c}\simeq \eta (1+\sigma _{0})$ is the coasting Lorentz factor. 
These scalings are based on the simplest first-order estimate. Since
photosphere emission would release some energy, $\Gamma _{\text{ra}}\neq $%
$\eta $; and synchrotron emission may also release some
energy, $\Gamma _{c}\neq \eta (1+\sigma _{0})$ \citep{Zhang2021}. $%
R_{\text{ra}}=\Gamma _{\text{ra}}r_{0}$ and $R_{c}=R_{\text{ra}}(\Gamma
_{c}/\Gamma _{\text{ra}})^{\mathbf{1/\delta }}$. The maximum $%
\delta $ of 1/3 is used in the following, except for Figure 9.

\subsection{\protect\bigskip The Photosphere Radius and the Comoving
Temperature}

With the dynamics in Equation $(\ref{b})$, the classical photosphere radius, 
$R_{\text{ph}}$, at which the scattering optical depth for a photon moving
in the radial direction drops down to unity ($\tau =1$), can be written as
(also see \citealt{Gao2015}) 
\begin{equation}
R_{\text{ph}}=\left\{ 
\begin{array}{l}
\left( \frac{L_{w}\sigma _{T}r_{0}^{2}}{8\pi m_{p}c^{3}\Gamma _{c}}\right)
^{1/3},\text{ \ \ \ \ \ \ \ \ }r_{0}<R_{\text{ph}}<R_{\text{ra}}; \\ 
\left( \frac{L_{w}\sigma _{T}R_{\text{ra}}^{2/3}}{8\pi m_{p}c^{3}\Gamma _{%
\text{ra}}^{2}\Gamma _{c}}\right) ^{3/5},\text{\ \ \ \ \ }R_{\text{ra}}<R_{%
\text{ph}}<R_{c}; \\ 
\frac{L_{w}\sigma _{T}}{8\pi m_{p}c^{3}\Gamma _{c}^{3}},\text{ \ \ \ \ \ \ \
\ \ \ \ \ \ \ \ }R_{\text{ph}}>R_{c}.%
\end{array}%
\right.
\end{equation}%
As shown in Figure 1, for the moderate magnetization $\sigma
_{0}\simeq 1-10$ and other parameter values adopted in this work, $%
R_{\text{ra}}<R_{\text{ph}}<R_{c}$ is almost satisfied. 

For a hybrid jet, the comoving temperature depends on whether there is
significant magnetic energy dissipation, namely magnetic energy is directly
converted to the heat, below the photosphere. Here, the photosphere emission
with significant magnetic dissipation %
\citep[e.g.][]{Thom1994,Ree2005,Gian2008} is not considered for the
following two reasons. On the one hand, for the hybrid jet with moderate
magnetization complete thermalization below the photosphere is likely to be
achieved (see section 4.1), thus the shape of the observed overall spectrum
mainly concerned on in this work is still the same as that for the
non-dissipative case. On the other hand, if complete thermalization could
not be achieved, the calculation is much too complicated and the rather high 
$E_{\text{p}}$ ( $\gtrsim $ 8 MeV, \citealt{Gian2006,Belo2013,Be2015})
predicted by the magnetically dissipative photosphere model is not
consistent with the observation.

Within non-dissipative case, the magnetic energy is only converted into the
kinetic energy of the bulk motion. This conversion may correspond to the
self-sustained magnetic bubbles or the helical jets %
\citep[e.g.][]{Spru2001,Uzde2006,Yuan2012}. Since the jet is
non-dissipative, adiabatic cooling with $r^{2}e^{3/4}\Gamma =$ const
proceeds \citep[e.g.][]{Pi1993}, here $e\propto T^{^{\prime }4}$. Thus,
considering the dynamical evolution in Equation $(\ref{b})$, the comoving
temperature is derived as 
\begin{equation}
T^{\prime }(r)=\left\{ 
\begin{array}{l}
T_{0}\left( \frac{r}{r_{0}}\right) ^{-1},\text{ \ \ \ \ \ \ \ \ \ \ \ \ \ \
\ \ \ }\ \ \ \ \ \ \ \ \ \ \ \ \ \ r_{0}<r<R_{\text{ra}}; \\ 
T_{0}\left( \frac{R_{\text{ra}}}{r_{0}}\right) ^{-1}\left( \frac{r}{R_{\text{%
ra}}}\right) ^{-\mathbf{(2+\delta )/3}},\text{\ \ \ \ \ \ \ \ \ \ }R_{\text{%
ra}}<r<R_{c}; \\ 
T_{0}\left( \frac{R_{\text{ra}}}{r_{0}}\right) ^{-1}\left( \frac{R_{c}}{R_{%
\text{ra}}}\right) ^{-\mathbf{(2+\delta )/3}}\left( \frac{r}{R_{c}}\right)
^{-2/3},\text{ \ \ \ \ }r>R_{c}.%
\end{array}%
\right.
\end{equation}%
Here, $T_{0}=\left[ L_{w}/4\pi r_{0}^{2}ac(1+\sigma _{0})\right] ^{1/4}$ is
the base outflow temperature at $r_{0}$, and $a$ is the radiation density
constant.

\subsection{Time-resolved Spectra from Probability Photosphere Emission}

\subsubsection{Impulsive Injection}

For the probability photosphere model, the observed spectrum is a
superposition of a series of blackbodies emitted from any place in the
outflow with a certain probability, which is calculated by 
\begin{equation}
F_{\nu }(\nu ,t)=\frac{N_{0}}{4\pi d_{L}^{2}}\iint P_{1}(r,\mu )P_{2}(\nu
,T)h\nu \times \delta (t-\frac{ru}{\beta c})d\mu dr\text{,}  \label{c}
\end{equation}%
where $u=1-\beta \cos \theta $, and $N_{0}=L_{w}/2.7k_{\text{B}}T_{0}$ is
the number of photons injected impulsively at the base of the outflow, $%
P_{1}(r,\mu )$ represents the probability density function for the final
scattering to occur at the coordinates $(r,\theta )$, $\mu =\cos \theta $, $%
P_{2}(\nu ,T)$ represents the probability for a photon of the observed
frequency $\nu $ last-scattered at $(r,\theta )$ with the observer frame
temperature of $T$. In following calculations, we adopt the two-dimensional
probability density function $P_{1}(r,\mu )$ as introduced in %
\citet{Belo2011}, i.e.,

\begin{eqnarray}
\frac{dP_{1}}{drd\mu } &=&D^{2}\frac{R_{\text{ph}}}{4r^{2}}\left\{ \frac{3}{2%
}+\frac{1}{\pi }\arctan [\frac{1}{3}(\frac{R_{\text{ph}}}{r}-\frac{r}{R_{%
\text{ ph}}})]\right\}  \notag \\
&&\times \exp \left[ -\frac{R_{\text{ph}}}{6r}(3+\frac{1-\mu ^{\prime }}{%
1+\mu ^{\prime }})\right] ,  \label{dd}
\end{eqnarray}%
where $\mu ^{\prime }=\cos \theta ^{\prime }$ is the value in the outflow
comoving frame and $D=[\Gamma (1-\beta \cos \theta )]^{-1}$ is the Doppler
factor.

\subsubsection{Continuous Wind}

It is more realistic to consider the continuous wind from the central engine
since the GRBs have relatively long duration ($> 1$ s). For simplicity, we
assume a constant $\eta =L_{w}(\hat{t})/\dot{M}(\hat{t})c^{2}$, here $\hat{t}
$ is the central-engine time since the earliest layer of the wind is
injected. According to the results in \citet{Deng2014}, the spectrum for the
case of constant wind luminosity without shut-down is similar to the
peak-flux spectrum for the case of variable wind luminosity, making the
simplicity reasonable.

For a layer ejected from $\hat{t}$ to $\hat{t}+d\hat{t}$, the observed
spectrum at the observer time $t$ is 
\begin{equation}
\hat{F}_{\nu }(\nu ,t,\hat{t})=\frac{N_{0}}{4\pi d_{L}^{2}}\iint P_{1}(r,\mu
)P_{2}(\nu ,T)h\nu \times \delta (t-\hat{t}-\frac{ru}{\beta c})d\mu dr\text{.%
}
\end{equation}%
Then, integrating over all the layers, we obtain the observed time-resolved
spectrum to be 
\begin{equation}
F_{\nu }(\nu ,t)=\int_{0}^{t}\hat{F}_{\nu }(\nu ,t,\hat{t})d\hat{t}\text{.}
\label{e}
\end{equation}

\section{Calculated Results}

\label{sec:result}

\subsection{Impulsive Injection}

\begin{figure*}
\centering\includegraphics[angle=0,height=7.3in]{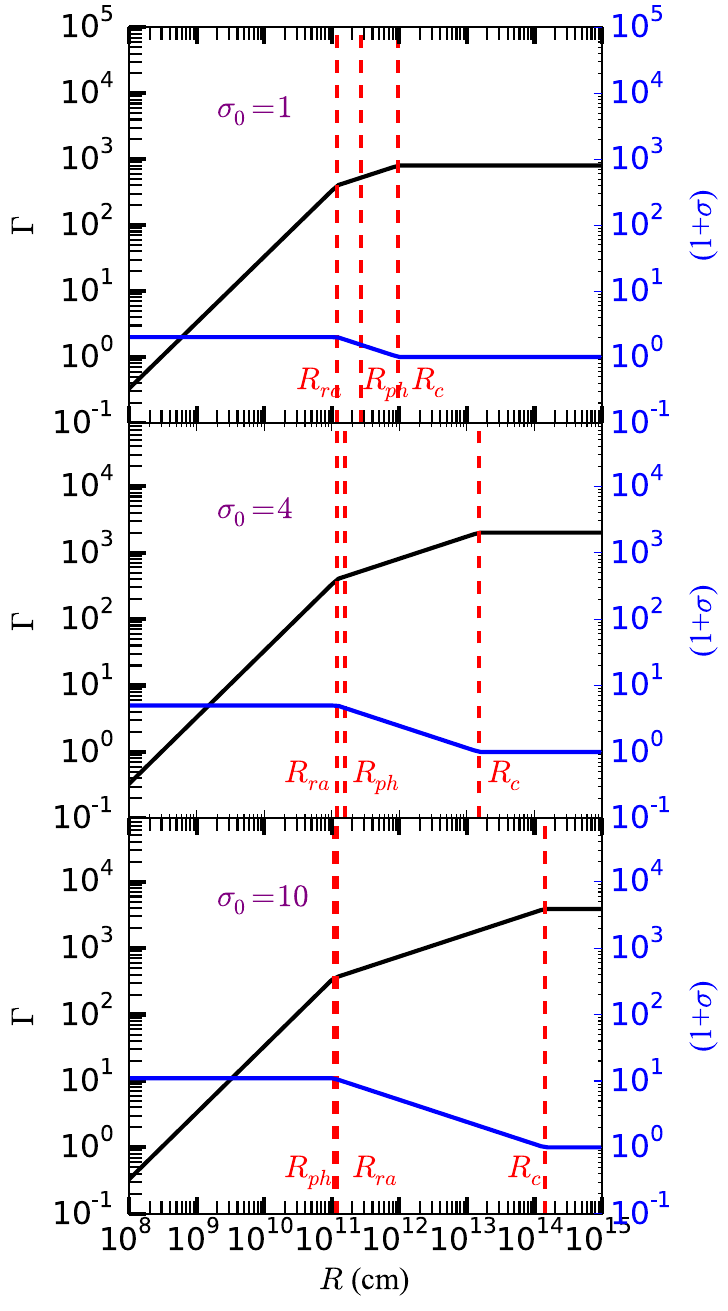} \ \ \ 
\caption{Jet dynamics and some characteristic radii for different
magnetization $\protect\sigma _{0}$. The black lines are for $%
\Gamma $ evolution and the blue lines are for $\protect\sigma $ evolution. Vertical red dashed lines denote three characteristic
radii: rapid acceleration radius $R_{\text{ra}}$, photosphere
radius $R_{\text{ph}}$, and coasting radius $R_{\text{c}}$%
. Following parameters are adopted: $L_{w}=10^{53}$ erg s$^{-1}$%
, $R_{0}=3\times 10^{8}$ cm, and $\protect\eta $ 
$=400$. Different panels correspond to different $\protect\sigma %
_{0}$: $\protect\sigma _{0}$ $=$ $1$
(top panel), $\protect\sigma _{0}$ $=$ $4$
(middle panel), and $\protect\sigma _{0}$ $=$ $10$ (bottom panel). For the moderate magnetization $\protect\sigma %
_{0}\simeq 1-10$ considered in this work, $R_{\text{ph}}$ 
$<$ $R_{\text{c}}$ is satisfied.}
\end{figure*}

\begin{figure*}
\centering
\includegraphics[angle=0,height=2.2in]{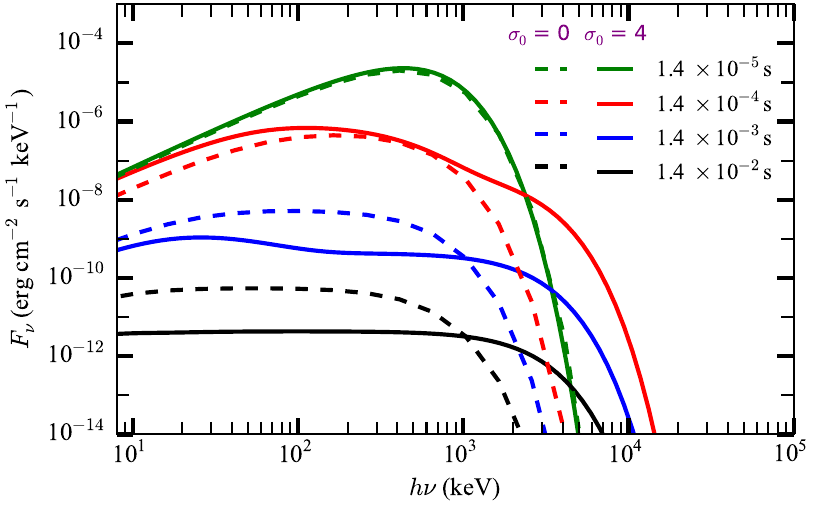} \ \ \ 
\caption{The time-resolved spectra of impulsive injection for the jet with
and without magnetization. The solid lines show the calculated time-resolved
spectra for the hybrid jet with a hot fireball component ($\protect\eta $ = $%
400$) and a cold Poynting-flux component ($\protect\sigma _{0}$= $4$). Also,
a total outflow luminosity of $L_{w}=10^{53}$ erg s$^{-1}$ is assumed,\ base
outflow radius $R_{0}=3\times 10^{8}$ cm, and luminosity distance $d_{\text{L%
}}=$ $4.85\times 10^{28}$ cm ($z=2$). Different colors represent different
observational times. For comparison, the dashed lines illustrate the
time-resolved spectra calculated in \citet{Deng2014} for the jet without
magnetization. Obviously, at later times the time-resolved spectra extend to
much higher energy and a power-law component emerges in the low-energy end.}
\label{Fig_1}
\end{figure*}

\begin{figure*}
\centering\includegraphics[angle=0,height=6.5in]{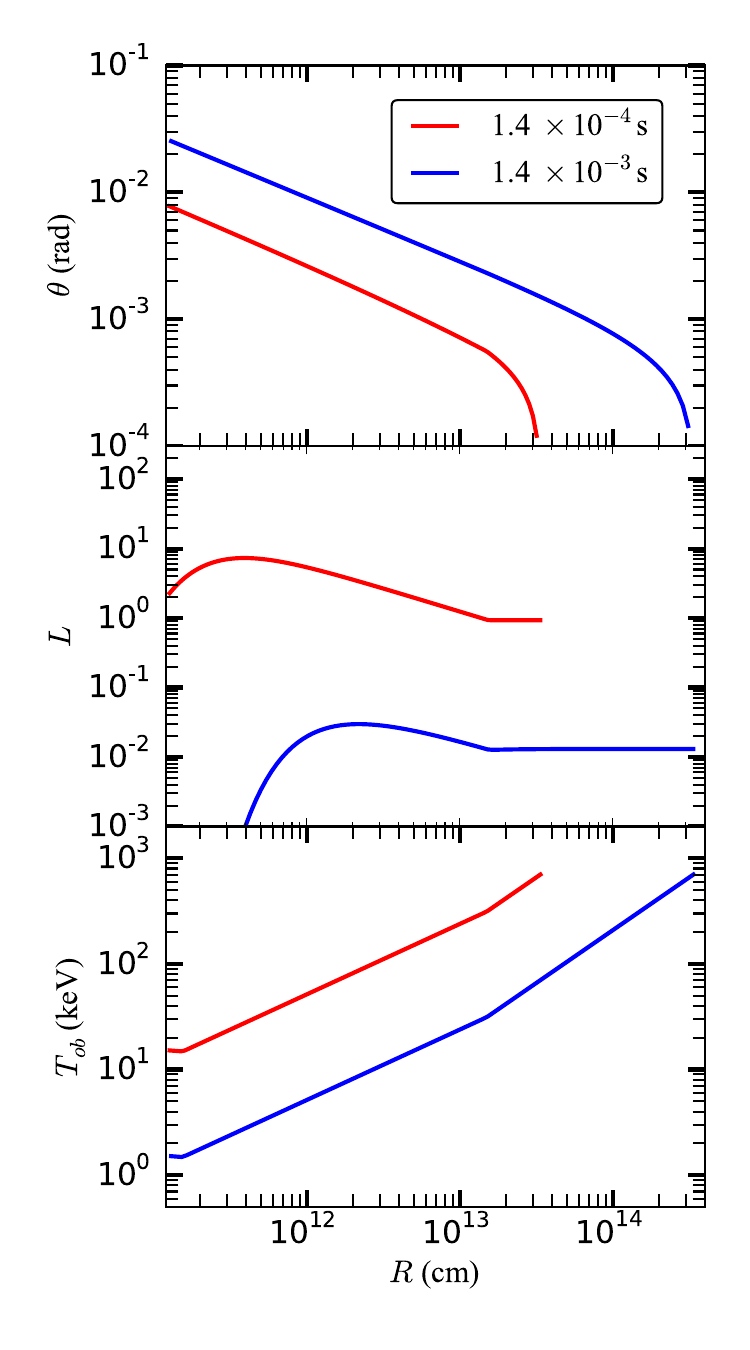} \ \ \ 
\caption{The detailed analysis of the formation of the time-resolved spectra
for impulsive injection by illustrating the emitted luminosity and the
observer-frame temperature at different positions of the equal arrival time
surface. Top panel: the distribution of the angle $\protect\theta $ and the
radius $r$ for the equal arrival time surface of different times, $1.4\times
10^{-4}$ s (red) and $1.4\times 10^{-3}$ s (blue). Middle panel: the emitted
luminosity in the corresponding radius $r$ for the equal arrival time
surface of different times. Bottom panel: the observer-frame temperature in
the corresponding radius $r$ for the equal arrival time surface of different
times.}
\label{Fig_2}
\end{figure*}

The time-resolved spectra of impulsive injection for the hybrid jet are
calculated by Equation $(\ref{c})$. \ Here we consider a moderate
magnetization $\sigma _{0}=$ $4$, since we try to explain the observed
spectra with only the thermal component while the non-thermal emission from
the Poynting flux is assumed to be weaker. Also, for this scenario the
observed relatively large radiative efficiency of the prompt emission $%
\epsilon _{\gamma }\equiv E_{\gamma }/(E_{\gamma }+E_{k})$ %
\citep{Lloy2004,Fan2006,Beni2016} constrains that the kinetic energy
powering the afterglow cannot be too large, thus the magnetization $\sigma
_{0}$ cannot be too large (for further discussion see section 4.2). %
Here we do not consider the magnetic dissipation. The magnetic dissipation
efficiency depends on the final $\sigma $ after the dissipation %
\citep{Deng2015}. Then, since the observed prompt luminosity is mainly
contributed by the thermal component, which is typically $\sim $ $10^{52}$
erg s$^{-1}$, we assume the total outflow luminosity to be $L_{w}=10^{53}$
erg s$^{-1}$. As for the base outflow radius $R_{0}$, we take $R_{0}=3\times
10^{8}$ cm close to the mean value of $10^{8.5}$ cm deduced in \citet{Pe2015}%
. The luminosity distance is assumed to be $d_{\text{L}}=$ $4.85\times
10^{28}$ cm, according to the peak of the GRB formation rate ($z=2$; see %
\citealt{Pes2016}). Thus, considering the redshift effect, to obtain the
observed typical peak energy $E_{\text{p}}\sim 300$ keV we take the
dimensionless entropy $\eta $ $=$ $400$.

The calculated time-resolved spectra with the above parameters are shown in
Figure 2 (the solid lines). Also, to illustrate the influence of the
magnetization, the time-resolved spectra calculated in \citet{Deng2014} for
the jet without magnetization are shown with the dashed lines. It is found
that at the later times (the high-latitude emission dominates) the
time-resolved spectra firstly show a power-law shape extending to a much
higher energy than the early-time blackbody. Then, the power-law component
vanishes gradually towards the low-energy end and the peak energy on the
high-energy end remains constant. To better understanding the formation of
the time-resolved spectra, in Figure 3 we show the distribution of the
emitted luminosity and the observer-frame temperature at different equal
arrival time surfaces (red lines for $1.4\times 10^{-4}$ s, and blue lines
for $1.4\times 10^{-3}$ s). For the emitted luminosity, we assume the number
of injected photon as $1$.

The emitted luminosity $L$ in the corresponding radius $r$ (with a
corresponding angle $\theta $, see the top panel) is shown in the middle
panel. While the calculation of this emitted luminosity $L$ is 
\begin{eqnarray}
L &=&\iint P_{1}(r,\mu )\delta (t-\frac{ru}{\beta c})d\mu dr  \notag \\
&=&\iint \text{ }D^{2}\frac{R_{\text{ph}}}{4r^{2}}\left\{ \frac{3}{2}+\frac{1%
}{\pi }\arctan [\frac{1}{3}(\frac{R_{\text{ph}}}{r}-\frac{r}{R_{\text{ph}}}%
)]\right\}  \notag \\
&&\times \exp \left[ -\frac{R_{\text{ph}}}{6r}(3+\frac{1-\mu ^{\prime }}{%
1+\mu ^{\prime }})\right] \times \delta (t-\frac{r(1-\beta \mu )}{\beta c}%
)d\mu dr  \notag \\
&=&\int [\Gamma (\frac{\beta ct}{r})]^{-2}\times \frac{R_{\text{ph}}}{4r^{2}}%
\left\{ \frac{3}{2}+\frac{1}{\pi }\arctan [\frac{1}{3}(\frac{R_{\text{ph}}}{r%
}-\frac{r}{R_{\text{ph}}})]\right\}  \notag \\
&&\times \exp \left[ -\frac{R_{\text{ph}}}{6r}(3+\frac{1-\mu ^{\prime }}{%
1+\mu ^{\prime }})\right] \times \frac{c}{r}dr\text{.}
\end{eqnarray}

For $r>R_{\text{ph}}$, we have 
\begin{eqnarray}
L &\simeq &[\Gamma (\frac{\beta ct}{r})]^{-2}\times \frac{R_{\text{ph}}}{%
4r^{2}}\times c  \notag \\
&\simeq &[\Gamma \beta ct]^{-2}\times \frac{R_{\text{ph}}}{4}\times c  \notag
\\
&\propto &\Gamma ^{-2}.
\end{eqnarray}%
Since the jet is still accelerated by the magnetic component as $\Gamma
\propto $ $r^{1/3}$ before the coasting radius $R_{c}$, for $R_{\text{ph}}$ $%
<r<R_{c}$ (see Figure 1) we obtain 
\begin{equation}
L\propto r^{-2/3}.
\end{equation}%
This correlation results in the power-law segment with a slope $\sim -2/3$
(both for $1.4\times 10^{-4}$ s and $1.4\times 10^{-3}$ s) in the middle
panel of Figure 3.

As for the observer-frame temperature $T^{\text{ob}}$, considering that the
comoving temperature $T^{\prime }(r)$ is constant for $R_{\text{ph}}$ $%
<r<R_{c}$, we have 
\begin{eqnarray}
T^{\text{ob}} &\equiv &D\cdot T^{\prime }(r)  \notag \\
&=&[\Gamma (r)\cdot (\frac{\beta ct}{r})]^{-1}\cdot T^{\prime }(r)  \notag \\
&\propto &r^{2/3}.
\end{eqnarray}%
Thus, a power-law component with a slope $\sim -1$ ($L\propto (T^{\text{ob}%
})^{-1}$, corresponding to $R_{\text{ph}}$ $<r<R_{c}$) shows up, for the
time-resolved spectra at the later times in Figure 2. Also, the peak energy
on the high-energy end for these spectra corresponds to the observer-frame
temperature at the line of sight ($\theta =0$) and the maximum radius $%
R_{\max }=\beta ct/(1-\beta )$, namely, $T^{\text{ob}}=\Gamma (R_{\max
})\cdot T^{\prime }(R_{\max })$. Since $R_{\max }>R_{\text{ph}}$, the peak
energy is surely much higher than that of the early-time blackbody as shown
in Figure 2, due to the larger $\Gamma (R_{\max })$ and the constant $%
T^{\prime }(R_{\max })$. For the much later times, with $R_{\max }>R_{c}$,
the peak energy remains unchanged because of the constant $\Gamma (R_{\max
}) $.

\subsection{Continuous Wind}

\begin{figure*}
\label{Fig_3} \centering\includegraphics[angle=0,height=2.05in]{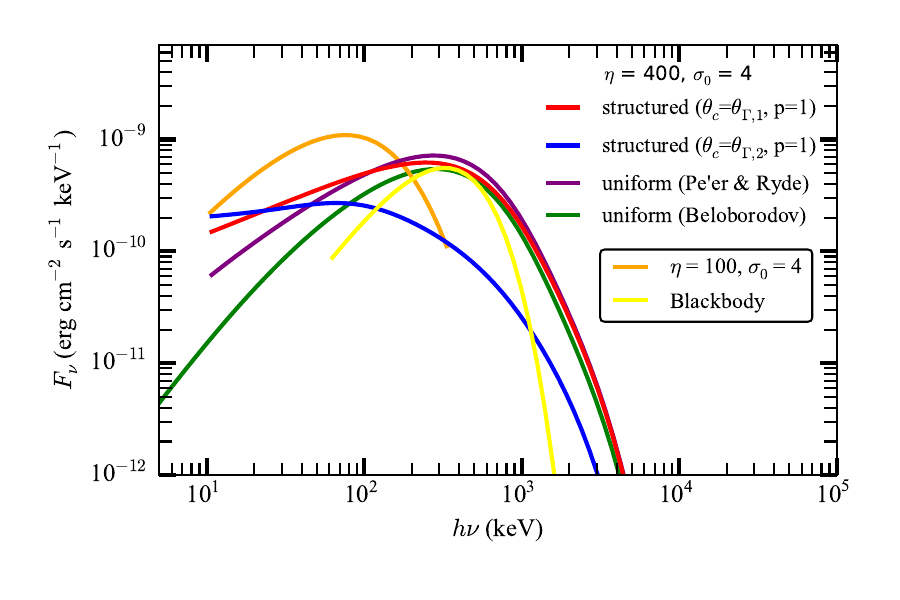} %
\centering\includegraphics[angle=0,height=2.05in]{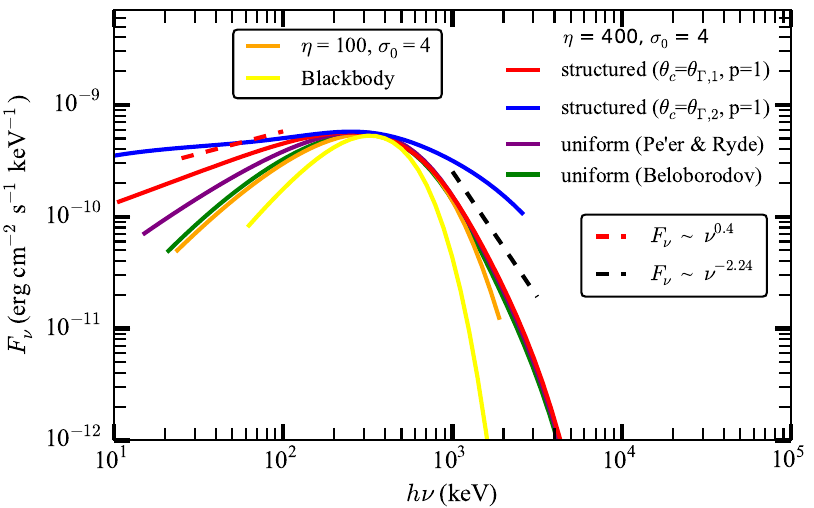} \ \ \ 
\caption{The time-resolved spectra at later times ($t$ $\geq $ $1$ s) of a
continuous wind for the hybrid jet with magnetization $\protect\sigma _{0}$= 
$4$ and various values or angular profiles of dimensionless entropy $\protect%
\eta $. Left panel: the calculated actual time-resolved spectra. Here, $%
\protect\theta_{\Gamma,1}=1/400$ and $\protect\theta_{\Gamma,2}=1/4000$.
Right panel: the time-resolved spectra which have been normalized to the
same peak energy and peak flux as the case of $\protect\eta $ = $400$ and $%
\protect\sigma _{0} $= $4$ (green line) to compare clearly the low-energy
and high-energy indices. For comparison, the yellow lines in the left and
right panels show the spectrum of blackbody. Also, the red and black dashed
lines in the right panel represent respectively the average low-energy index
($-0.6$) and high-energy index ($-3.24$) fitted with the Band function for
the time-resolved spectra of a large sample of single pulses in 
\citet{Yu2018}.}
\end{figure*}

\begin{figure*}
\label{Fig_4} \centering\includegraphics[angle=0,height=2.2in]{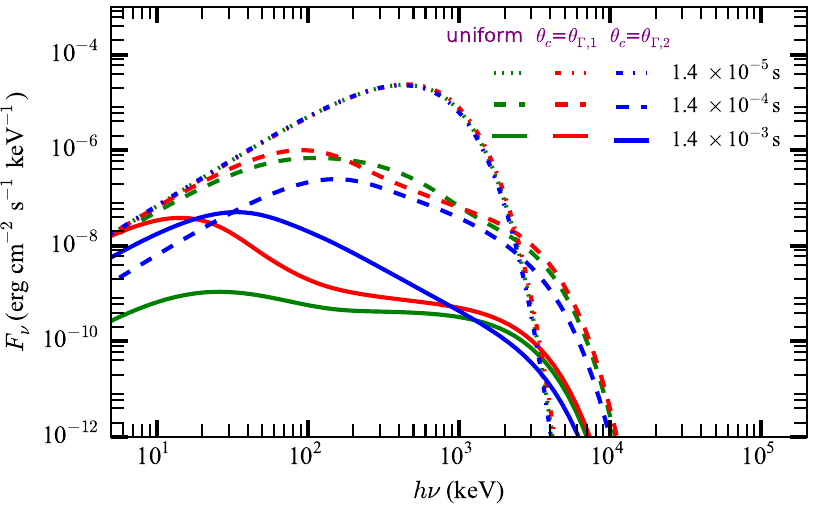} \ \ \ 
\caption{Comparison of the time-resolved spectra of impulsive injection for
the hybrid jet with different angular profiles of dimensionless entropy $%
\protect\eta $. The green lines are the time-resolved spectra in Figure 2
for a uniform profile. While the red and blue lines show the time-resolved
spectra for the angular profile with an inner constant core (width of the
core $\protect\theta _{c}=\protect\theta_{\Gamma,1}$ for the red lines and $%
\protect\theta_{c}=\protect\theta_{\Gamma,2}$ for the blue lines) and outer
power-law decreased component (power-law index $p=1$). Also, the dashed and
solid lines are for the time-resolved spectra at $1.4\times 10^{-4}$ s and $%
1.4\times 10^{-3}$ s, respectively.}
\end{figure*}

\begin{figure*}
\label{Fig_5} \centering\includegraphics[angle=0,height=2.15in]{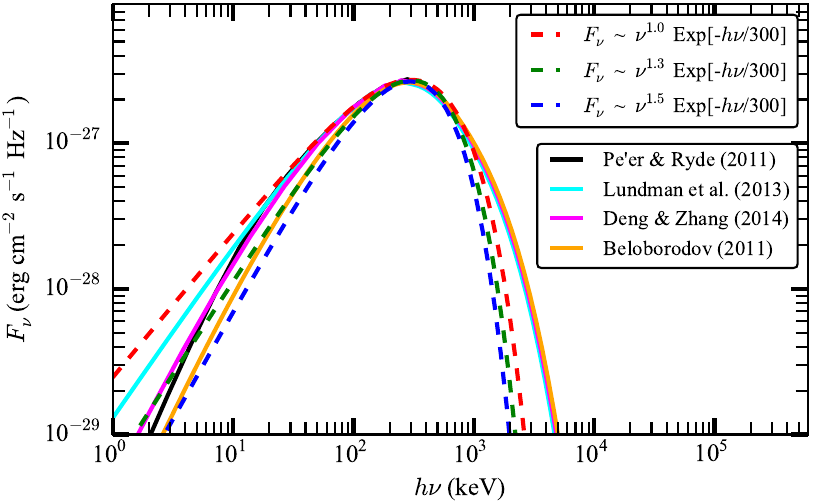} \ \ \ 
\caption{Comparison of the calculated time-resolved spectra at later times ($%
t$ $\geq $ $1$ s) of a continuous wind with different probability density
functions. The parameters of the jet are the same as those used in 
\citet{Deng2014}. The three dashed lines show the spectra (normalized to the
same peak energy and peak flux as the calculated spectra) of the cutoff
power-law model with the low-energy photon index of $0$ (red), $0.3$ (green)
and $0.5$ (blue), respectively.}
\end{figure*}

For the continuous wind with constant wind luminosity, the spectrum at a
later time ($t$ $\geq $ $1$ s) which corresponds to the observed peak-flux
spectrum is calculated by Equation $(\ref{e})$. This calculated
time-resolved spectrum, with the same parameters as Figure $2$ ($\eta $ = $%
400$, $\sigma _{0}$= $4$) is illustrated by the green lines (with the
probability density function introduced in \citealt{Belo2011}) in Figure $4$%
. Apparently, the spectrum on the high-energy end is a power law rather than
an exponential cutoff, which is the natural result of the power-law
component extending to much higher energy in Figure $2$. The purple lines
are calculated by the probability density function described in %
\citet{Pe2011}. Also, the spectrum for a smaller dimensionless entropy $\eta 
$ ($\eta $ = $100$) is shown by the orange line in Figure $4$, which is
found to have a smaller peak energy (see the left panel). This is consistent
with the positive correlation $T^{\text{ob}}\propto \eta ^{16/15}$ in
Equation (25) of \citet{Gao2015} for the regime of $\eta >(1+\sigma
_{0})^{1/2}$ and $R_{\text{ra}}<R_{\text{ph}}<R_{c}$, the regime mainly
considered in our work. It is worth noting that, the peak energy weakly
depends on the other parameters, i.e., $\sigma _{0}$, $L_{w}$ and $R_{0}$.

Furthermore, we calculate the spectrum (with the probability density
function described in \citealt{Pe2011}) for the case of the dimensionless
entropy $\eta $ of a lateral structure, with which the observed typical
low-energy spectral index $\alpha \approx -1$ can be obtained for the
unmagnetized jet under the framework of the probability photosphere model %
\citep{Lund2013,Meng2019}. The considered angular Lorentz
factor profile consists of an inner constant core with width $\theta _{c}$
and the outer power-law decreased component with power-law index $p$. 
Note that an angle-independent luminosity is considered, because the
spectrum expected to be observed is formed by the photons making their last
scattering at approximately $\lesssim 5/$ $\Gamma _{0}$ 
(as shown in \citealt{Lund2013}) and the isotropic angular width for %
Lorentz factor $\theta _{c,\Gamma }$ is likely to be much smaller
than the isotropic angular width for luminosity $\theta _{c,L}$ 
(see the top panels of Figures 8 and 9 in \citealt{ZhaWoo2003}, see also
section 4.3 for more discussion). The red lines in Figure $4$ represent the
calculated spectrum for a hybrid jet with $\theta _{c}=\theta _{\Gamma
,1}=1/400$ and $p=1$, corresponding to the typical value $\alpha \sim -1$
for an unmagnetized jet. While the blue lines are the spectrum for hybrid
jet with $\theta _{c}=\theta _{\Gamma ,2}=1/4000$ and $p=1$, corresponding
to the minimum value $\alpha \sim -2$ for unmagnetized jet (see Figure 7d in %
\citealt{Meng2019}). To compare clearly the low-energy and high-energy
indices, in Figure $4$ the calculated actual time-resolved spectra (left
panel) have been normalized to the same peak energy and peak flux as the
case of $\eta $ = $400$ and $\sigma _{0}$= $4$ (shown in the right panel).

We then find that the spectrum of the hybrid jet possesses the much harder
low-energy spectral index than that of the unmagnetized jet with the same
angular profile, $\alpha \sim -0.6$ for $\theta _{c}=\theta_{\Gamma,1}$, $%
p=1 $ and $\alpha \sim -1$ for $\theta_{c}=\theta_{\Gamma,2}$, $p=1$. To
better understand the origin of this hardness, in Figure $5$ we show the
time-resolved spectra of impulsive injection for these two angular profiles.
The power-law segment caused by the magnetic acceleration, similar to that
for the uniform profile in Figure $2$ and noteworthily with the swallower
slope ($\sim -1$) than that of the power law ($\sim -2$) for the
unmagnetized structured jet \citep{Meng2019}, exists close to and below the
peak energy of the early-time blackbody in the late-time spectra. This
results in the above hardness for the spectrum of continuous wind. On the
high-energy end, with the angular profile of the dimensionless entropy, the
spectrum remains to be a power law. The high-energy power-law index $\beta $
is the same as the uniform case for $\theta_{c}=\theta_{\Gamma,1}$, $p=1$ ($%
\beta \sim -4$, see Figure $9$ and the discussion in Section $3.3$), and
much larger ($\beta \sim -1.6$) for $\theta_{c}=\theta_{\Gamma,2}$, $p=1$.
In a word, the spectrum of the hybrid jet is analogue to the empirical Band
function \citep{Band1993} spectrum, whereas the spectrum of the unmagnetized
jet corresponds to the spectrum of the empirical cutoff power-law model %
\citep{Meng2019}, within the framework of the probability photosphere model.

Interestingly, in all the statistical works of a large sample of GRBs %
\citep[e.g.][]{Kan2006,Gold2012,Gold2013,Grub2014,Yu2016}, the average
low-energy spectral index for the GRBs best-fitted by the Band function is
harder ($0.1\sim 0.3$) than that for the GRBs best-fitted by the cutoff
power-law model, for both the time-integrated and the peak-flux spectra.
This hardness is quite consistent with our results for the probability
photosphere model discussed in the previous paragraph. Also, for the
distributions of the low-energy and high-energy indices, our results
(corresponding to the peak-flux spectra) are quite similar to the
statistical results of the peak-flux spectra for the GRBs best-fitted by the
Band function. For the low-energy spectral index, the typical value is $%
\alpha \sim -0.6$ and the minimum value $\alpha \sim -1$. While for the
high-energy spectral index, the maximum value $\beta \sim -1.6$ in our model
is close to the statistical result. The typical value $\beta \sim -4$ in our
model seems to be much softer than the well-known $\beta \sim -2.5$, which
we consider to arise from the assumption of the single pulse, namely without
the overlap of the pulses. This consideration is proposed recently in %
\citet{Yu2018}, where the time-resolved spectral analysis of a large sample
of single pulses has been performed and the average high-energy spectral
index ($\beta \sim -3.24$) indeed is found to be much softer. Noteworthily,
the average high-energy spectral index $\beta \sim -3.24$ in that work can
be well reproduced with our hybrid jet model, by considering the parameter
dependence on the power-law index of magnetic acceleration $\delta $ (see
Figure $9$ and the discussion in Section $3.3$). In the right panel of
Figure $4$, the red and black dashed lines represent the average low-energy
index ($-0.6$) and high-energy index ($-3.24$) in that work, respectively.

In the following, we give a detailed discussion on the maximum (or the
hardest) low-energy spectral index $\alpha $. The maximum $\alpha $ from the
statistical works is $\alpha \sim 0$, for the GRBs best-fitted by the Band
function and the cutoff power-law model both. For the probability
photosphere model, with magnetization or not, the maximum $\alpha $
corresponds to a uniform jet. The low-energy index $\alpha $ of the
calculated spectrum for a uniform jet is regarded as $\alpha \sim 0.5$ in %
\citet{Deng2014}, while $\alpha \sim 0$ is obtained in \citet{Lund2013} (see
the high-energy inner jet spectral component for a wide jet $\theta _{\text{j%
}}=10/\Gamma _{0}$ and $\theta _{\text{v}}=0$ in Figure $8$, the red
diamonds and solid black lines) and \citet{Meng2019} (see Figure $7b$
therein). We notice that, in \citet{Deng2014} (see Figure $3$ and Figure $9$%
) $\alpha \sim 0.5$ is taken from the power-law index close to $3$ keV. But
this is quite rough, since the actual power-law index is determined by the
fitting of the observed spectrum from $8$ keV to the peak energy $\sim $ $%
300 $ keV, quite above $3$ keV. Thus, in Figure $6$ we compare the
calculated spectra in \citet{Deng2014} (orange and magenta lines) with the
spectra of the cutoff power-law model for $\alpha =$ $0$, $\alpha =0.3$ and $%
\alpha =0.5 $. The orange line is calculated by the probability density
function introduced in \citet{Belo2011}, while the magenta line is for the
probability density function proposed in \citet{Deng2014}. Notice that, the
spectra of the cutoff power-law model have been normalized to the same peak
energy and peak flux as the calculated spectra. In Figure $6$, we also plot
the calculated spectra with the probability density function described in %
\citet{Pe2011} and \citet{Lund2013}, using the same jet parameters as those
in \citet{Deng2014}. It is found that, except for the probability density
function in \citet{Belo2011}, the calculated spectra for the other three
kinds of probability density function are similar, all close to the cutoff
power-law spectrum for $\alpha =$ $0$ from $\sim $ $20$ keV to the peak
energy $\sim $ $300$ keV. This is consistent with the maximum $\alpha $ from
the statistical works. While the calculated spectrum for the probability
density function in \citet{Belo2011} is quite close to the cutoff power-law
spectrum for $\alpha =$ $0.3$. We think two aspects are responsible for this 
hardness: the angle corresponding to the observer-frame temperature
of $\sim $ $8$ keV to $\sim $ $300$ keV is small ($\lesssim 5/\Gamma _{0}$,
also see the top and bottom panels in Figure $3$); and the probability
density function for the small angle in \citet{Belo2011} is not as good as
the others (see the solid lines in Figure 7 of \citealt{Deng2014}).

\subsection{Parameter Dependence}

\begin{figure*}
\label{Fig_6} \centering\includegraphics[angle=0,height=2.05in]{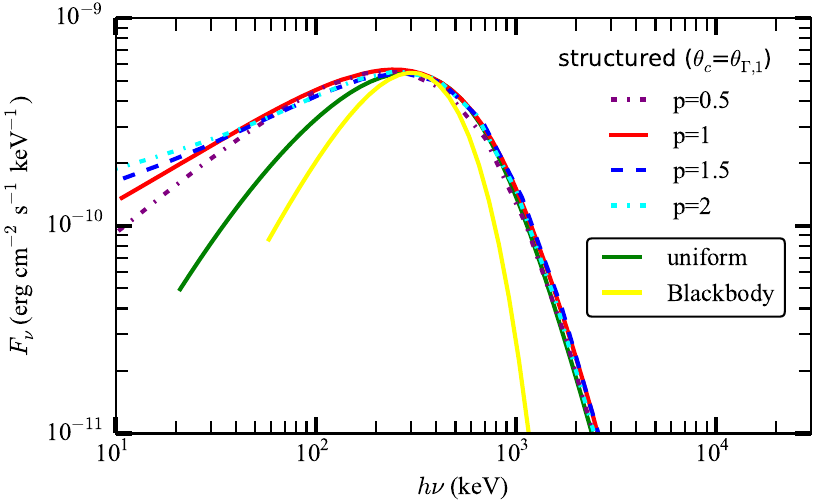} %
\centering\includegraphics[angle=0,height=2.05in]{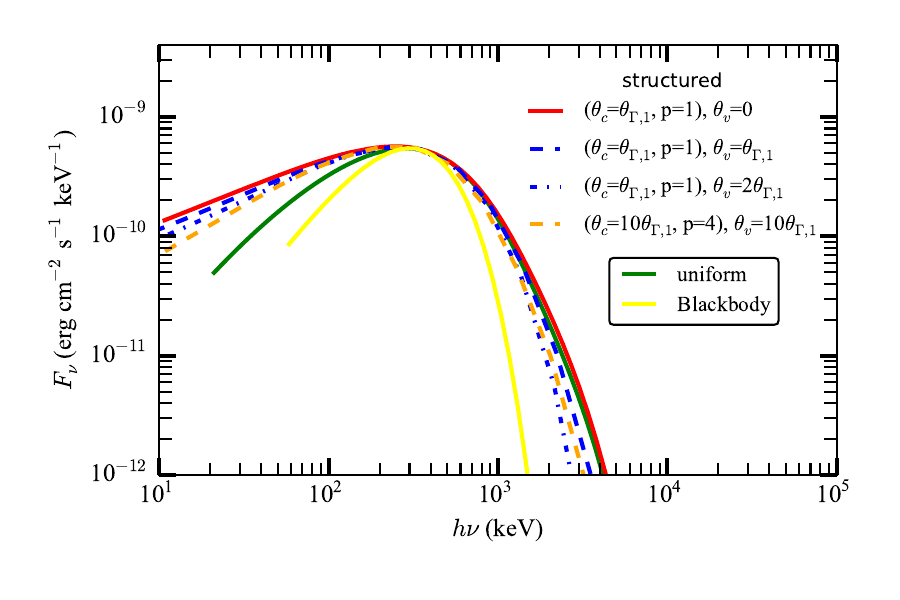} \ \ \ 
\caption{The calculated time-resolved spectra at later times ($t$ $\geq $ $1$
s) of continuous wind for a structural angular profiles of dimensionless
entropy $\protect\eta $ with different power-law index $p$ or non-zero
viewing angle. Left panel: the dependence of the time-resolved spectra on
the power-law index, $p=1$ (red solid), $p=0.5$ (magenta dash-dotted), $%
p=1.5 $ (blue dashed) and $p=2$ (cyan dash-dotted). Right panel: the
time-resolved spectra with non-zero viewing angle. The dashed and
dash-dotted lines are for $\protect\theta_{v}=\protect\theta_{\Gamma,1}$ and 
$\protect\theta_{v}=2\cdot \protect\theta_{\Gamma,1}$ respectively, along
with $\protect\theta_{c}=\protect\theta_{\Gamma,1}$ and $p=1$. While the
orange dashed line is for $\protect\theta _{v}=10\cdot \protect\theta%
_{\Gamma,1}$ with $\protect\theta_{c}=10\cdot \protect\theta_{\Gamma,1}$ and 
$p=4$. The green and yellow lines are the same as those in Figure 4. }
\end{figure*}
\ \ \ 

\begin{figure*}
\label{Fig_7} \centering\includegraphics[angle=0,height=2.05in]{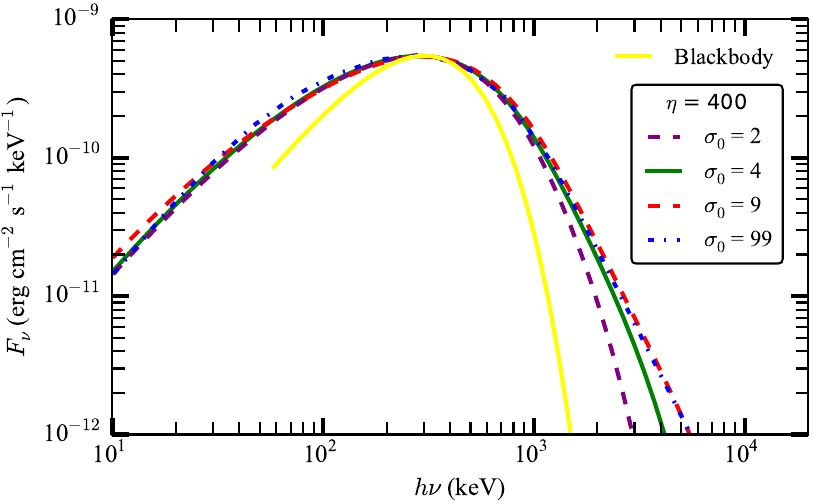} %
\centering\includegraphics[angle=0,height=2.05in]{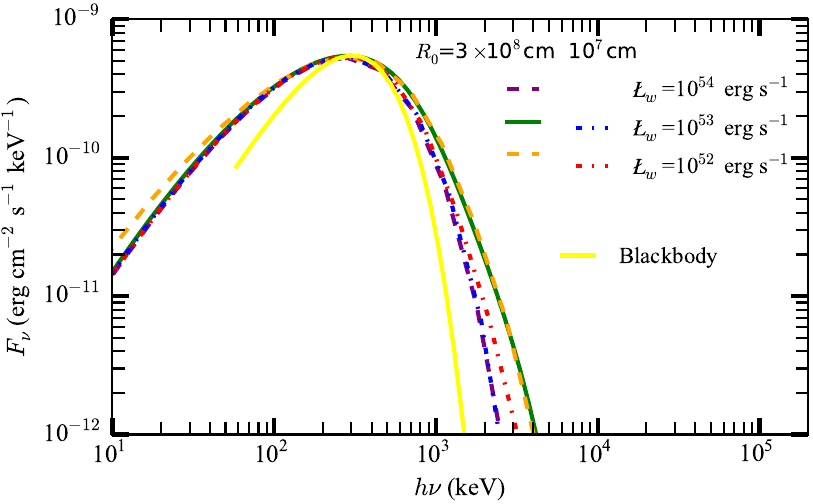} \ \ \ 
\caption{The calculated time-resolved spectra at later times ($t$ $\geq $ $1$
s) of continuous wind with different magnetization $\protect\sigma _{0}$ or
combination of $L_{w}$ and $R_{0}$. Left panel: the dependence of the
time-resolved spectra on the magnetization, $\protect\sigma _{0}=2$ (purple
dashed), $\protect\sigma _{0}=9$ (red dashed) and $\protect\sigma _{0}=99$
(blue dash-dotted). Right panel: the time-resolved spectra with different
combination of $L_{w}$ and $R_{0}$. The purple and orange dashed lines are
for $L_{w}=10^{54}$ erg s$^{-1}$ and $L_{w}=10^{52}$ erg s$^{-1}$
respectively, along with $R_{0}=3\times 10^{8}$ cm. While the blue and red
dash-dotted lines are for $L_{w}=10^{53}$ erg s$^{-1}$ and $L_{w}=10^{52}$
erg s$^{-1}$ respectively, along with $R_{0}=10^{7}$ cm. The green and
yellow lines are also the same as those in Figure 4. }
\end{figure*}

\begin{figure*}
\label{Fig_8} \centering\includegraphics[angle=0,height=2.05in]{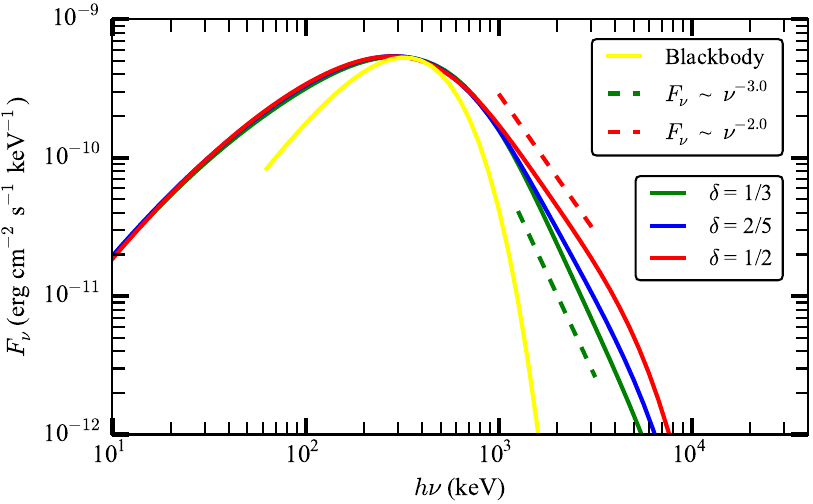} \ \ \ 
\caption{The calculated time-resolved spectra at later times ($t$ $\geq $ $1$
s) of continuous wind with different power-law index of magnetic
acceleration $\protect\delta $. The green, blue and red lines are for $%
\protect\delta =1/3$, $\protect\delta =2/5$ and $\protect\delta =1/2$
respectively, along with magnetization $\protect\sigma _{0}=9$. While the
green and red dashed lines represent the high-energy spectral index $\protect%
\beta =-4$ and $\protect\beta =-3$, respectively. The yellow line is the
spectrum of blackbody. }
\end{figure*}

As shown in Figure 4, the calculated spectrum of continuous wind for our
hybrid jet model can reproduce the observed low-energy and high-energy
indices quite well, for the GRBs best-fitted by the Band function. In this
section, we analyze in detail the dependence of these spectral indices on
the power-law index of the dimensionless entropy profile $p$ and the viewing
angle $\theta _{v}$ (Figure 7), the magnetization $\sigma _{0}$ and the
combination of $L_{w}$ and $R_{0}$ (Figure 8), and the power-law index of
magnetic acceleration $\delta $ (Figure 9). In Figures 7-9, all the
calculated spectra have been normalized to the same peak energy and peak
flux as the case of $\eta $ = $400$ and $\sigma _{0}$= $4$ (green line).

From the left panel of Figure 7 we can see that, with $\theta_{c}=1/400$ and
different values of $p$, the high-energy spectrum is almost the same and the
low-energy spectral index does not change a lot. The low-energy spectral
index is slightly softer for $p=1.5$ and $p=2$, while slightly harder for $%
p=0.5$. Also, the influence of non-zero viewing angle on the spectrum is
shown in the right panel of Figure 7. With a non-zero viewing angle $\theta
_{v}=\theta_{\Gamma,1}$ or $\theta _{v}=2\cdot \theta_{\Gamma,1}$ the
spectrum becomes narrower, namely, the low-energy spectrum is harder and the
high-energy spectrum deviates a little from the power law to the exponential
cutoff. This is consistent with the spectral result of smaller $\eta $ (see
the orange line in the right panel of Figure 4) and quite different from the
unmagnetized case, in which the shape of the spectrum remains unchanged with
a non-zero viewing angle. In addition, the low-energy spectral index for $%
\theta _{c}=10\cdot \theta_{\Gamma,1}$, $p=1$ and $\theta_{v}=10\cdot
\theta_{\Gamma,1}$ is found to be much harder than that of the unmagnetized
case ($\alpha $ $\sim $ $-1$), just as the situation for $%
\theta_{c}=\theta_{\Gamma,1}$, $p=1$ mentioned above.

The dependence of the shape of the calculated spectrum on the magnetization $%
\sigma _{0}$ is illustrated in the left panel of Figure 8. Notice that we
take the total luminosity $L_{w}=6\times 10^{52}$ erg s$^{-1}$ for $\sigma
_{0}=2$, to have comparable prompt emission luminosity with the case of $%
\sigma _{0}=4$. And we keep $L_{w}=10^{53}$ erg s$^{-1}$ for $\sigma _{0}=9$
and the extreme case of $\sigma _{0}=99$, to account for the $R_{\text{ph}%
}<R_{\text{ra}}$ regime. With smaller magnetization $\sigma _{0}$ ($\sigma
_{0}=2$), the spectrum is narrower in both the low-energy (slightly) and
high-energy (significantly) ends. If the magnetization is larger ($\sigma
_{0}=9$ or even $\sigma _{0}=99$), entering the $R_{\text{ph}}<R_{\text{ra}}$
regime, the low-energy spectrum is a little softer and the high-energy power
law can extend to much higher energy with the approximate slope as the $%
\sigma _{0}=4$ case. This softness and higher energy is due to the larger
range of $R_{\text{ph}}$ $<r<R_{c}$, which is responsible for the
high-energy power law with negative index (see Figure 3 and the discussion
there).

In the right panel of Figure 8, we plot the calculated spectra for different
combinations of $L_{w}$ and $R_{0}$. The low-energy spectrum is almost the
same except for the case of $L_{w}=10^{52}$ erg s$^{-1}$ and $R_{0}=3\times
10^{8}$ cm, which enters the $R_{\text{ph}}<R_{\text{ra}}$ regime and thus
has the slightly softer low-energy spectrum. The shape of the high-energy
spectrum depends on the comparison of $R_{\text{ph}}$ and $R_{\text{ra}}$.
For $L_{w}=10^{52}$ erg s$^{-1}$ and $R_{0}=3\times 10^{8}$ cm, the
high-energy spectrum is close to the reference spectrum. For the other three
combinations, since $R_{\text{ph}}$ is much greater than $R_{\text{ra}}$,
the high-energy spectrum becomes narrower.

Note that in the above discussion, for the magnetically driven acceleration,
we only consider the case of magnetic reconnection for the non-axisymmetric
rotator \citep{Dren2002,DrenSpru2002}. Whereas, for an initially
axisymmetric flow, the acceleration can happen through the kink instability %
\citep{Lyu2003,GianSpru2006}. Besides, the magnetic acceleration driven by
the kink instability seems to be more rapid than the magnetic reconnection
case, with a larger power-law index $\delta $ of $\sim 2/5$ or even $\sim
1/2 $ (see Figure 5 in \citealt{GianSpru2006}). Thus, in Figure 9 we compare
the calculated spectra of our model for $\delta =1/3$, $\delta =2/5$ and $%
\delta =1/2$, and find that their high-energy power-law indices are quite
different while the low-energy indices remain the same. For $\delta =1/3$,
we have $\beta =-4$; for $\delta =1/2$, we obtain $\beta =-3$; and $\beta $
lies between $-4$ and $-3$ for $\delta =2/5$. Surprisingly, this rough
distribution of $\beta $ ($-4$ to $-3$) is well consistent with the $\beta $
distribution of the softer cluster for a large sample of single pulses in %
\citet{Yu2018}, where the $\beta $ distribution seems to show the bimodal
distribution with a harder typical cluster (peaks at $-2.5$) and a softer
cluster (peaks at $-3.5$; see Figure 1 therein).

\section{Discussion}

\label{sec:dis}

\subsection{Magnetic Dissipation}

Significant magnetic dissipation may happen below the photosphere according
to several previous works %
\citep[e.g.][]{Thom1994,Ree2005,Gian2008,Me2011,Vere2012}, thus enhancing
the photosphere emission. However, in this sub-photosphere region complete
thermalization may not be achieved due to the lack of enough photons (low
creation rate), for the case of Poynting-flux-dominated outflow. Then, the
photosphere spectrum could have a non-thermal shape with an ultra-high peak
energy $E_{p}$ ranging from 1 MeV to about 20 MeV %
\citep{Gian2006,Belo2013,Be2015}. But we think that, for the hybrid jet with
moderate magnetization considered in this work, complete thermalization may
be achieved because of the existence of the extra thermal component in the
outflow. Thus, the spectrum emitted at a particular position could be a
blackbody with the temperature a bit larger than the non-dissipative case
(see Equation (30) in \citealt{Gao2015}). The shape of the observed overall
spectrum is still the same as that for the non-dissipative case.

\subsection{Radiative Efficiency}

The radiative efficiency of the prompt emission $\epsilon _{\gamma }$,
generally defined as $E_{\gamma }/(E_{\gamma }+E_{k})$, is a crucial
quantity to distinguish different prompt emission models. Here, $E_{\gamma }$
means the radiated energy in the prompt phase and $E_{k}$ means the
remaining kinetic energy in the afterglow phase. For the photosphere
emission model of a hybrid jet with moderate magnetization considered in
this work, we have

\begin{eqnarray}
\epsilon _{\gamma } &=&\frac{E_{\gamma }}{E_{\gamma }+E_{k}}  \notag \\
&=&\frac{L_{\text{Th}}\times (R_{\text{ph}}/R_{\text{ra}})^{-7/9}}{L_{\text{%
Th}}+L_{P}}  \notag \\
&=&\frac{(R_{\text{ph}}/R_{\text{ra}})^{-7/9}}{1+\sigma _{0}}\text{,}
\end{eqnarray}%
in the typical $R_{\text{ph}}\geq R_{\text{ra}}$ regime. While, $\epsilon
_{\gamma }=1/(1+\sigma _{0})$ is obtained in the $R_{\text{ph}}\leq R_{\text{%
ra}}$ regime. Note that the magnetic energy is thought to be transferred to
the kinetic energy completely before\ the onset of the afterglow, because of
the moderate magnetization. Then, for $\sigma _{0}\sim 2-9$ considered
above, $\epsilon _{\gamma }$ ranges from a few percents to $\sim 0.3$.
Namely, our photosphere model predicts a relatively low efficiency of $\sim
0.1$.

Observationally, $E_{k}$ can be inferred through the late-time X-ray
afterglow \citep{Kuma2000,Free2001}. Then, along with the obtained $%
E_{\gamma }$ by integrating the prompt spectrum, the radiative efficiency $%
\epsilon _{\gamma }$ can be obtained. The inferred radiative efficiency is
quite high, $\epsilon _{\gamma }\sim $ $0.4$ $-1.0$, in most of previous
studies \citep{Lloy2004,Berg2007,Nyse2009,DAva2012,Wygo2016}. This favors
greatly the photosphere emission model without magnetization, especially
with larger $R_{0}$ inferred in \citet{Pe2015} since $R_{\text{ph}}$ is more
close to saturation radius $R_{\text{s}}$. Note that the above method
considers that the late-time X-ray afterglow is contributed by fast cooling
electrons (for the typical values of magnetic equipartition parameter $%
\epsilon _{B}\simeq 0.01-0.1$) . If $\epsilon _{B}$ is smaller ( $\sim
10^{-4}$) for a portion of GRBs as\ inferred in several works %
\citep{Barn2014,Santa2014,Wang2015,ZhaBB2015}, slow cooling or significant
Inverse Compton losses take place and the estimated radiative efficiency $%
\epsilon _{\gamma }\sim 0.1$ is smaller \citep{Fan2006,Beni2016}. This
portion of GRBs may correspond to the photosphere emission with moderate
magnetization discussed in this work.

\subsection{Availability of the Assumption of  $\protect\theta %
_{c,\Gamma }<\protect\theta _{c,L}$}

As stated above, since the observed spectrum is contributed by the
photons emitted from a rather narrow angular region ($\lesssim 5/$  
 $\Gamma _{0}$) and $\theta _{c,\Gamma }<\theta _{c,L}$ is
likely to be the real situation based on the simulation, we take $%
dL/d\Omega \approx $ const in our calculations. Namely, we only
consider the case of smaller viewing angle ($\theta _{v}<\theta _{c,L}$%
). The assumption of $\theta _{c,\Gamma }<\theta _{c,L}$
is supported in some prior simulations, including both hydrodynamical ones %
\citep{Lazza2006,Ito2021} and magnetohydrodynamical ones (MHD, %
\citealt{Tche2008,Geng2019}).

Theoretically, $\theta _{c,\Gamma }<\theta _{c,L}$ could
be understood as a natural result of the enhanced material density for
larger angle (see Figure 3 in \citealt{Lazza2006}). The structured jet is
produced because the jet will be collimated by the progenitor envelope (or
dynamical ejecta) when penetrating it. This progenitor envelope (or
dynamical ejecta) is matter-dominated, and makes the shocked jet have an
increased material density (for larger angle) when collimation happens.
Then, since $\Gamma (\theta )\propto L/M(\theta )$, the Lorentz
factor will start to decrease even when the $L$ remains constant.
Namely, $\theta _{c,\Gamma }<\theta _{c,L}$ is obtained.

However, due to the complexity of simulations and lack of robust
results, $\theta _{c,\Gamma }$ $>$ $\theta _{c,L}$ may also be common in realistic situations. Here, in Figure 10 we
also perform calculations for the case of $\theta _{c,\Gamma }=\theta
_{c,L} $, which is often adopted for structured jet in the
literature. The other parameters are the same as those in Figure 4. Besides, 
$\theta _{c,L}=\theta _{c,\Gamma }=1/400$, and q is the power-law decreased
index for luminosity. In this case, the low-energy spectral index $\alpha $ 
is much harder. This is because the low-energy photons emitted from
the high-latitude region become much less (since the luminosity is smaller).
Then, the observed bursts with harder low-energy spectral index can be
reproduced in this case. Furthermore, according to recent studies %
\citep{Meng2018,Burgess2020} $\alpha $ may not be a good indicator
to justify the radiation mechanism, we should use the model
spectrum to directly fit the data.

\begin{figure*}
\label{Fig_10} \centering\includegraphics[angle=0,height=2.05in]{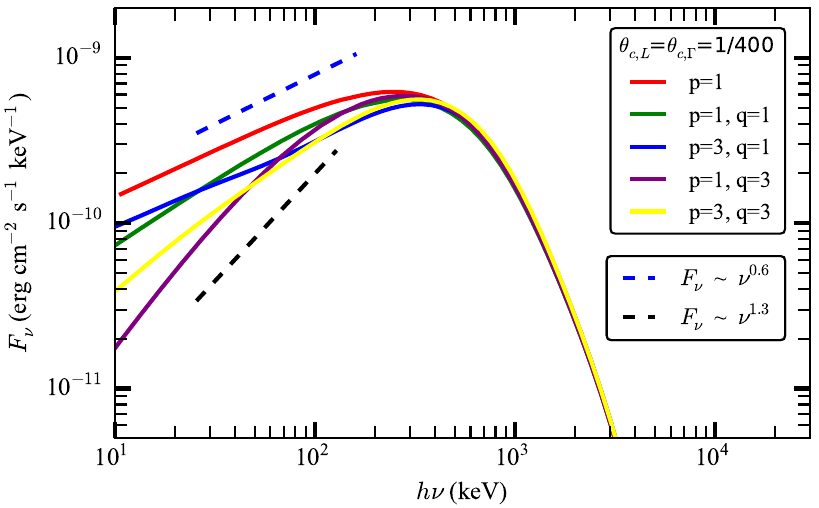}
\ \ \ 
\caption{The time-resolved spectra at later times ($t$ $%
\geq $ $1$ s) of a continuous wind for structured jet
with $\protect\theta _{c,L}=\protect\theta _{c,\Gamma }=1/400$ and
various power-law decreased indices of luminosity (q) and Lorentz factor
(p). A hybrid jet with magnetization $\protect\sigma _{0}$ = $4$  is considered, and the Lorentz factor in the isotropic core is $%
400$. The red line is the same as that in Figure 4, for an
angle-independent luminosity ($\protect\alpha \sim -0.6$%
). Obviously, for the combinations of p=1 and q=1 (the green line)
and of p=3 and q=1 (the blue line), the low-energy spectral index $\protect%
\alpha $ is only slightly harder than that of the angle-independent
luminosity case.\ While for the combinations of p=1 and q=3 (the purple
line) and of p=3 and q=3 (the yellow line), the low-energy spectral index $%
\protect\alpha $ is much harder. The blue and black dashed lines
represent the low-energy spectral indices of $-0.4$ ($F_{\protect%
\nu } \sim \protect\nu ^{0.6}$) and $0.3$ ($F_{\protect\nu %
} \sim \protect\nu ^{1.3}$), respectively.}
\end{figure*}

\subsection{Impact on the observed spectra by the synchrotron
emission}

Synchrotron emission may also contribute to the observed spectrum
for a hybrid jet. Considering the synchrotron emission from magnetic
dissipation, the high-energy spectrum may be less steeper when peak energy
of the photosphere spectrum is comparable with that of the synchrotron
spectrum. Otherwise, a two-peak spectrum is likely to exist. For smaller $%
\sigma _{0}$, the photosphere component may dominant, while
synchrotron component may dominant for larger $\sigma _{0}$.

\section{CONCLUSIONS}

\label{sec:conclu}

In this paper, by invoking the probability photosphere model we investigate
the shape of the photospheric emission spectrum for the hybrid outflow,
which contains a thermal component and a magnetic component with moderate
magnetization ($\sigma _{0}=L_{P}/L_{\text{Th}}\sim 1-10$). The following
conclusions are drawn.

(1) The photosphere spectrum on the high-energy end is a power law rather
than an exponential cutoff. This high-energy power-law component arises from
the continued increase of the bulk Lorentz factor $\Gamma $ (due to the
magnetically driven acceleration of the magnetic component) and the constant
comoving temperature $T^{\prime }$ above the photosphere radius $R_{\text{ph}%
}$, where the emission is not negligible (though less) according to the
probability photosphere model. The power-law segment can extend to higher
energy with larger magnetization $\sigma _{0}$ (smaller $L_{w}$, larger $%
R_{0}$, or larger $\eta $), because of the larger range of $R_{\text{ph}}$ $%
<r<R_{c}$ (responsible for the high-energy power law).

(2) With the similar angular profiles of the dimensionless entropy $\eta $
as the unmagnetized jet, considered in previous works %
\citep{Lund2013,Meng2019}, the distribution of the low-energy indices
(corresponding to the peak-flux spectra) for our photosphere model is quite
consistent with the statistical result of the peak-flux spectra for the GRBs
best-fitted by the Band function. For a combination of $\theta _{c}=1/\Gamma
=1/400$ and $p=1$, with the unmagnetized probability photosphere model, the
resulted $\alpha $ is consistent with the\ observed typical value $\alpha
\sim -1.0$ for the GRBs best-fitted by the cutoff power law. Considering the
magnetized probability photosphere model in this work, the obtained $\alpha $
is accordant with the\ observed typical value $\alpha \sim -0.6$ for the
GRBs best-fitted by the Band function. While for a combination of an
extremely narrow core $\theta _{c}=1/(10\Gamma )=1/4000$ and $p=1$, with the
unmagnetized probability photosphere model, the resulted $\alpha $ is
consistent with the\ observed minimum value $\alpha \sim -2.0$ for the GRBs
best-fitted by the cutoff power law. Considering the magnetized probability
photosphere model in this work, the obtained $\alpha $ is accordant with
the\ observed minimum value $\alpha \sim -1$ for the GRBs best-fitted by the
Band function. Also, by analyzing the low-energy\ spectra\ for uniform jet\
calculated with\ different probability density functions, we find that the
hardest $\alpha $ predicted by the probability photosphere model (both
unmagnetized and magnetized) should be $\alpha \sim 0$, almost the same as
both the observed maximum values for the GRBs best-fitted by the cutoff
power law and the Band function.\ 

(3) The high-energy power-law index $\beta $ for our photosphere model
solely depends on the power-law index of magnetic acceleration $\delta $, if
only the core for the angular profile of $\eta $ is not too narrow. After
considering the magnetic acceleration due to magnetic reconnection for the
non-axisymmetric rotator ($\delta \sim 1/3$) and kink instability in an
initially axisymmetric flow ($\delta \sim 1/2$), the distribution of the
obtained $\beta $ (from $-4$ to $-3$; $\beta =-4$ for $\delta =1/3$, and $%
\beta =$ $-3$ for $\delta =1/2$) is well consistent with the $\beta $
distribution of the softer cluster for a large sample of single pulses in %
\citet{Yu2018}.\ Besides, for an extremely narrow core $\theta
_{c}=1/(10\Gamma )=1/4000$ and $p=1$, much larger $\beta \sim $ $-1.6$ is
obtained, similar to the maximum value of the statistical result. In total,
the observed $\beta $ distribution could be well interpreted with the
photosphere model in this work.

\section*{Acknowledgements}

We thank the anonymous referee for constructive suggestions. This work is supported by the National Natural Science Foundation of China (Grant Nos. 11725314, 12041306, 11903019, 11833003), the Major Science and
Technology Project of Qinghai Province (2019-ZJ-A10). Y.Z.M. is supported by
the National Postdoctoral Program for Innovative Talents (grant No.
BX20200164).

\section*{Data availability}

The data underlying this article will be shared on reasonable request to the
corresponding author.

\label{lastpage}


\begin{thebibliography}{Veres \& M{\'e}sz{\'a}ros(2012)}
\bibitem[Abdo et al.(2009)]{Abdo2009} Abdo A.~A., et al., 2009, \apjl, 706, L138

\bibitem[Abramowicz et al.(1991)]{Abra1991} Abramowicz M.~A., Novikov I.~D., Paczynski B., 1991, \apj, 369, 175

\bibitem[Acuner \& Ryde(2018)]{Acun2018} Acuner Z., Ryde F., 2018, \mnras, 475, 1708

\bibitem[Acuner et al.(2020)]{Acun2020} Acuner Z., Ryde F., Pe'er A., Mortlock D., Ahlgren B., 2020, \apj, 893, 128

\bibitem[Aloy et al.(2005)]{Aloy2005} Aloy M.~A., Janka H.-T., M{\"u}ller E., 2005, \aap, 436, 273

\bibitem[Axelsson et al.(2012)]{Axel2012} Axelsson M., et al., 2012, \apjl, 757, L31

\bibitem[Axelsson \& Borgonovo(2015)]{AxBo2015} Axelsson M., Borgonovo L., 2015, \mnras, 447, 3150

\bibitem[Band et al.(1993)]{Band1993} Band D., et al., 1993, \apj, 413, 281

\bibitem[Barniol Duran(2014)]{Barn2014} Barniol Duran R., 2014, \mnras,
442, 3147

\bibitem[B{\'e}gu{\'e} \& Pe'er(2015)]{Be2015} B{\'e}gu{\'e} D., Pe'er A.\ 2015, \apj, 802, 134

\bibitem[Beloborodov(2011)]{Belo2011} Beloborodov A.~M., 2011, \apj, 737, 68

\bibitem[Beloborodov(2013)]{Belo2013} Beloborodov A.~M., 2013, \apj, 764,
157

\bibitem[Beloborodov(2017)]{Belo2016} Beloborodov A.~M., 2017, \apj, 838,
125

\bibitem[Beniamini et al.(2016)]{Beni2016} Beniamini P., Nava L., Piran T., 2016, \mnras, 461, 51

\bibitem[Berger(2007)]{Berg2007} Berger E., 2007, \apj, 670, 1254

\bibitem[Burgess et al.(2017)]{Bur2017} Burgess J.~M., Greiner J., B{\'e}gu{\'e} D.,  Berlato F., 2017, arXiv:1710.08362

\bibitem[Burgess(2019)]{Burgess2019} Burgess J.~M., 2019, \aap,
629, A69

\bibitem[Burgess et al.(2020)]{Burgess2020} Burgess J.~M., B{\'e}gu{\'e} D., Greiner J., Giannios D., Bacelj A., Berlato F., 2020, Nature Astronomy, 4, 174

\bibitem[Dai \& Gou(2001)]{Dai2001} Dai Z.~G., Gou L.~J., 2001, \apj,
552, 72

\bibitem[D'Avanzo et al.(2012)]{DAva2012} D'Avanzo P., et al., 2012, \mnras, 425, 506

\bibitem[Deng \& Zhang(2014)]{Deng2014} Deng W., Zhang B., 2014, \apj,
785, 112

\bibitem[Deng et al.(2015)]{Deng2015} Deng W., Li H., Zhang B., Li S., 2015, \apj, 805, 163

\bibitem[Drenkhahn(2002)]{Dren2002} Drenkhahn G., 2002, \aap, 387, 714

\bibitem[Drenkhahn \& Spruit(2002)]{DrenSpru2002} Drenkhahn G., Spruit H.~C., 2002, \aap, 391, 1141

\bibitem[Duan, \& Wang(2019a)]{Duan2019} Duan M.-Y., Wang X.-G., 2019, \apj, 884, 61

\bibitem[Duan, \& Wang(2019b)]{Duan2019b} Duan M.-Y., Wang X.-G., 2020, ApJ, 890, 90

\bibitem[Fan \& Piran(2006)]{Fan2006} Fan Y.-Z., Piran T., 2006, \mnras,
369, 197

\bibitem[Fan et al.(2012)]{Fan2012} Fan Y.-Z., Wei D.-M., Zhang F.-W., Zhang B.-B., 2012, \apjl, 755, L6

\bibitem[Ford et al.(1995)]{Ford1995} Ford L.~A., et al., 1995, \apj, 439, 307

\bibitem[Freedman \& Waxman(2001)]{Free2001} Freedman D.~L., Waxman E., 2001, \apj, 547, 922

\bibitem[Gao \& Zhang(2015)]{Gao2015} Gao H., Zhang B., 2015, \apj,
801, 103

\bibitem[Geng et al.(2018)]{Geng18} Geng J.-J., Huang Y.-F., Wu X.-F., Zhang B., Zong H.-S., 2018, \apjs, 234, 3

\bibitem[Geng et al.(2019)]{Geng2019} Geng J.-J., Zhang B., K{\"o}lligan A., Kuiper R., Huang Y.-F., 2019, \apjl, 877, L40

\bibitem[Ghirlanda et al.(2010)]{Ghir2010} Ghirlanda G., Nava L., Ghisellini G., 2010, \aap, 511, A43

\bibitem[Ghirlanda et al.(2013)]{Ghir2013} Ghirlanda G., Pescalli A., Ghisellini G., 2013, \mnras, 432, 3237

\bibitem[Giannios(2006)]{Gian2006} Giannios D., 2006, \aap, 457, 763

\bibitem[Giannios \& Spruit(2006)]{GianSpru2006} Giannios D., Spruit H.~C., 2006, \aap, 450, 887

\bibitem[Giannios(2008)]{Gian2008} Giannios D., 2008, \aap, 480, 305

\bibitem[Goldstein et al.(2012)]{Gold2012} Goldstein A., et al.,2012, \apjs, 199, 19

\bibitem[Goldstein et al.(2013)]{Gold2013} Goldstein A., Preece R.~D., Mallozzi R.~S., Briggs M.~S., Fishman G.~J., Kouveliotou C., Paciesas W.~S.,Burgess J.~M., 2013, \apjs, 208, 21

\bibitem[Goodman(1986)]{Good1986} Goodman J., 1986, \apjl, 308, L47

\bibitem[Gottlieb et al.(2021)]{Gott2020} Gottlieb O., Nakar E., Bromberg O., 2021, MNRAS, 500, 3511

\bibitem[Granot et al.(2011)]{Gran2011} Granot J., Komissarov S.~S., Spitkovsky A., 2011, \mnras, 411, 1323

\bibitem[Gruber et al.(2014)]{Grub2014} Gruber D., et al., 2014, \apjs, 211, 12

\bibitem[Guiriec et al.(2013)]{Gui2013} Guiriec S., et al., 2013, \apj, 770, 32

\bibitem[Guiriec et al.(2011)]{Gui2011} Guiriec S., et al., 2011, \apjl, 727, L33

\bibitem[Hou et al.(2018)]{Hou2018} Hou S.-J., et al., 2018, \apj, 866, 13

\bibitem[Huang et al.(2019)]{Huang2019} Huang B.-Q., Lin D.-B., Liu T., Ren J., Wang X.-G., Liu H.-B., Liang E.-W., 2019, \mnras, 487, 3214

\bibitem[Ito et al.(2021)]{Ito2021} Ito H., Just O., Takei Y., Nagataki S., 2021, ApJ, 918, 59

\bibitem[Kaneko et al.(2006)]{Kan2006} Kaneko Y., Preece R.~D., Briggs M.~S., Paciesas W.~S., Meegan C.~A., Band D.~L., 2006, \apjs, 166, 298

\bibitem[Komissarov et al.(2009)]{Komi2009} Komissarov S.~S., Vlahakis N., K{\"o}nigl A., Barkov M.~V., 2009, \mnras, 394, 1182

\bibitem[Komissarov et al.(2010)]{Komi2010} Komissarov S.~S., Vlahakis N., K{\"o}nigl A., 2010, \mnras, 407, 17

\bibitem[Kumar(2000)]{Kuma2000} Kumar P., 2000, \apjl, 538, L125

\bibitem[Kumar \& Granot(2003)]{Kumar2003} Kumar P., Granot J., 2003, %
\apj, 591, 1075

\bibitem[Larsson et al.(2015)]{Lar2015} Larsson J., Racusin J.~L., Burgess J.~M., 2015, \apjl, 800, L34

\bibitem[Lazzati et al.(2007)]{Lazza2006} Lazzati D., Morsony B.~J., 
Begelman M.~C., 2007, Philosophical Transactions of the Royal Society of London Series A, 365, 1141

\bibitem[Lazzati et al.(2013)]{Lazz2013} Lazzati D., Morsony B.~J.,
Margutti R., Begelman M.~C., 2013, \apj, 765, 103

\bibitem[Lei et al.(2013)]{Lei2013} Lei W.-H., Zhang B., Liang E.-W.,
2013, \apj, 765, 125

\bibitem[Li(2019a)]{Li2019a} Li L., 2019a, \apjs, 242, 16

\bibitem[Li et al.(2019b)]{Li2019b} Li L., et al., 2019b, \apj, 884, 109

\bibitem[Li(2019c)]{Li2019c} Li L., 2019c, \apjs, 245, 7

\bibitem[Li(2019d)]{Li2019d} Li L., 2020, ApJ, 894, 100

\bibitem[Liang \& Kargatis(1996)]{Lia1996} Liang E., Kargatis V., 1996, \nat, 381, 49

\bibitem[Lin et al.(2018)]{Lin2018} Lin D.-B., Liu T., Lin J., Wang X.-G., Gu W.-M., Liang E.-W., 2018, \apj, 856, 90

\bibitem[Lloyd-Ronning \& Zhang(2004)]{Lloy2004} Lloyd-Ronning N.~M., Zhang B., 2004, \apj, 613, 477

\bibitem[Lu et al.(2010)]{Lu2010} Lu R.-J., Hou S.-J., Liang E.-W., 2010, \apj, 720, 1146

\bibitem[Lu et al.(2012)]{Lu2012} Lu R.-J., Wei J.-J., Liang E.-W., Zhang B.-B., L{\"u} H.-J., L{\"u} L.-Z., Lei W.-H., Zhang B., 2012, \apj, 756, 112

\bibitem[Lu et al.(2017)]{Lu2017} Lu R.-J., Du S.-S., Cheng J.-G., L{\"u} H.-J., Zhang H.-M., Lan L., Liang E.-W., 2017, arXiv:1710.06979

\bibitem[Lundman et al.(2013)]{Lund2013} Lundman C., Pe'er A., Ryde F., 2013, \mnras, 428, 2430

\bibitem[Lyutikov \& Blandford(2003)]{Lyu2003} Lyutikov M., Blandford R., 2003, arXiv:astro-ph/0312347

\bibitem[MacFadyen \& Woosley(1999)]{Mac1999} MacFadyen A.~I., Woosley S.~E., 1999, \apj, 524, 262

\bibitem[Meng et al.(2018)]{Meng2018} Meng Y.-Z., et al., 2018, \apj, 860, 72

\bibitem[Meng et al.(2019)]{Meng2019} Meng Y.-Z., Liu L.-D., Wei J.-J., Wu X.-F., Zhang B.-B., 2019, \apj, 882, 26

\bibitem[M{\'e}sz{\'a}ros \& Rees(2000)]{Me2000} M{\'e}sz{\'a}ros P., Rees M.~J., 2000, \apj, 530, 292

\bibitem[M{\'e}sz{\'a}ros(2002)]{Me2002} M{\'e}sz{\'a}ros P., 2002, \araa, 40, 137

\bibitem[M{\'e}sz{\'a}ros \& Rees(2011)]{Me2011} M{\'e}sz{\'a}ros P., Rees M.~J., 2011, \apjl, 733, L40

\bibitem[Metzger et al.(2011)]{Met2011} Metzger B.~D., Giannios D., Thompson T.~A., Bucciantini N., Quataert E., 2011, \mnras, 413, 2031

\bibitem[Mizuta, Nagataki \& Aoi(2011)]{Mizu2011} Mizuta A., Nagataki S., Aoi J., 2011, \apj, 732, 26

\bibitem[Morsony, Lazzati \& Begelman(2007)]{Mor2007} Morsony B.~J., Lazzati D., Begelman M.~C., 2007, \apj, 665, 569

\bibitem[Murguia-Berthier et al.(2017)]{Murgu2017} Murguia-Berthier A., et al., 2017, ApJL, 835, L34

\bibitem[Nagakura et al.(2011)]{Naga2011} Nagakura H., Ito H., Kiuchi K., Yamada S., 2011, ApJ, 731, 80

\bibitem[Nysewander et al.(2009)]{Nyse2009} Nysewander M., Fruchter A.~S., Pe'er A., 2009, \apj, 701, 824

\bibitem[Paczynski(1986)]{Pac1986} Paczynski B., 1986, \apjl, 308, L43

\bibitem[Pe'er(2008)]{Pe2008} Pe'er A., 2008, \apj, 682, 463

\bibitem[Pe'er \& Ryde(2011)]{Pe2011} Pe'er A., Ryde F., 2011, \apj,
732, 49

\bibitem[Pe'er et al.(2015)]{Pe2015} Pe'er A., Barlow H., O'Mahony S., Margutti R., Ryde F., Larsson J., Lazzati D., Livio M., 2015, \apj, 813, 127

\bibitem[Pescalli et al.(2016)]{Pes2016} Pescalli A., et al., 2016, \aap, 587, A40

\bibitem[Piran et al.(1993)]{Pi1993} Piran T., Shemi A., Narayan R., 1993, \mnras, 263, 861

\bibitem[Piran(1999)]{Pi1999} Piran T., 1999, \physrep, 314, 575

\bibitem[Preece et al.(2000)]{Pree2000} Preece R.~D., Briggs M.~S., Mallozzi R.~S., Pendleton G.~N., Paciesas W.~S., Band D.~L., 2000, \apjs, 126, 19

\bibitem[Rees \& Meszaros(1994)]{Ree1994} Rees M.~J., Meszaros, P., 1994, \apjl, 430, L93

\bibitem[Rees \& M{\'e}sz{\'a}ros(2005)]{Ree2005} Rees M.~J., M{\'e}sz{\'a}ros P., 2005, \apj, 628, 847

\bibitem[Rossi et al.(2002)]{Rossi2002} Rossi E., Lazzati D., Rees M.~J., 2002, \mnras, 332, 945

\bibitem[Rosswog(2013)]{Ross2013} Rosswog S., 2013, Philosophical
Transactions of the Royal Society of London Series A, 371, 20120272

\bibitem[Ruffini et al.(2013)]{Ru2013} Ruffini R., Siutsou I.~A., Vereshchagin G.~V., 2013, \apj, 772, 11

\bibitem[Ryde(2004)]{Ry2004} Ryde F., 2004, \apj, 614, 827

\bibitem[Ryde(2005)]{Ry2005} Ryde F., 2005, \apjl, 625, L95

\bibitem[Ryde \& Pe'er(2009)]{Ry2009} Ryde F., Pe'er A., 2009, \apj,
702, 1211

\bibitem[Ryde et al.(2010)]{Ry2010} Ryde F., et al., 2010, \apjl, 709, L172

\bibitem[Ryde et al.(2017)]{Ry2017} Ryde F., Lundman C., Acuner Z., 2017, \mnras, 472, 1897

\bibitem[Santana et al.(2014)]{Santa2014} Santana R., Barniol Duran R., Kumar P., 2014, \apj, 785, 29

\bibitem[Sapountzis \& Vlahakis(2014)]{Sap2014} Sapountzis K., Vlahakis N., 2014, Physics of Plasmas, 21, 072124

\bibitem[Spruit et al.(2001)]{Spru2001} Spruit H.~C., Daigne F., Drenkhahn G., 2001, \aap, 369, 694

\bibitem[Tchekhovskoy et al.(2008)]{Tche2008} Tchekhovskoy A., McKinney J.~C., Narayan R., 2008, \mnras, 388, 551

\bibitem[Thompson(1994)]{Thom1994} Thompson C., 1994, \mnras, 270, 480

\bibitem[Uzdensky \& MacFadyen(2006)]{Uzde2006} Uzdensky D.~A., MacFadyen A.~I., 2006, \apj, 647, 1192

\bibitem[Veres \& M{\'e}sz{\'a}ros(2012)]{Vere2012} Veres P., M{\'e}sz{\'a}ros P., 2012, \apj, 755, 12

\bibitem[Vurm \& Beloborodov(2016)]{Vur2016} Vurm I., Beloborodov A.~M., 2016, \apj, 831, 175

\bibitem[Wang et al.(2020)]{Wang2020} Wang K., et al., 2020, ApJ, 899, 111

\bibitem[Wang et al.(2015)]{Wang2015} Wang X.-G., et al., 2015, \apjs, 219, 9

\bibitem[Wang et al.(2021)]{Wang2021} Wang X.~I., et al., 2021, arXiv:2107.10452

\bibitem[Wygoda et al.(2016)]{Wygo2016} Wygoda N., Guetta D., Mandich M.~A., Waxman E., 2016, \apj, 824, 127

\bibitem[Yang \& Zhang(2018)]{Yang2018} Yang Y.-P., Zhang B., 2018, \apjl, 864, L16

\bibitem[Yang et al.(2020)]{Yang2020} Yang J., et al., 2020, \apj, 899, 106

\bibitem[Yu et al.(2015)]{Yu2015} Yu H.-F., van Eerten H.~J., Greiner J., Sari R., Narayana Bhat P., von Kienlin A., Paciesas W.~S., Preece, R. D., 2015, \aap, 583, A129

\bibitem[Yu et al.(2016)]{Yu2016} Yu H.-F., et al., 2016, \aap, 588, A135

\bibitem[Yu et al.(2019)]{Yu2018} Yu H.-F., Dereli-B{\'e}gu{\'e} H., Ryde F., 2019, \apj, 886, 20

\bibitem[Yuan \& Zhang(2012)]{Yuan2012} Yuan F., Zhang B., 2012, \apj,
757, 56

\bibitem[Zhang \& M{\'e}sz{\'a}ros(2002)]{ZhMe2002} Zhang B., M{\'e}sz{\'a}ros P., 2002a, \apj, 571, 876

\bibitem[Zhang et al.(2007)]{ZhaB2007} Zhang B., et al., 2007, \apj, 655, 989

\bibitem[Zhang(2011)]{ZhaB2011} Zhang B., 2011, Comptes Rendus Physique,
12, 206

\bibitem[Zhang \& Yan(2011)]{ZhaYan2011} Zhang B., Yan H., 2011, \apj,
726, 90

\bibitem[Zhang(2020)]{ZhaB2020} Zhang B., 2020, Nature Astronomy, 4, 210

\bibitem[Zhang et al.(2021)]{Zhang2021} Zhang B., Wang Y., Li L., 2021, \apjl, 909, L3

\bibitem[Zhang \& Zhang(2014)]{ZhaBo2014} Zhang B., Zhang B., 2014, \apj%
, 782, 92

\bibitem[Zhang et al.(2011)]{ZhaBB2011} Zhang B.-B., et al., 2011, \apj, 730, 141

\bibitem[Zhang et al.(2015)]{ZhaBB2015} Zhang B.-B., van Eerten H., Burrows D.~N., Ryan G.~S., Evans P.~A., Racusin J.~L., Troja E., MacFadyen A., 2015, \apj, 806, 15

\bibitem[Zhang et al.(2016)]{Zhang2016} Zhang B.-B., Uhm Z.~L., Connaughton V., Briggs M.~S., Zhang B., 2016,\apj,816,72

\bibitem[Zhang et al.(2018a)]{ZhaBB2018} Zhang B.-B., et al., 2018a, Nature Astronomy, 2, 69

\bibitem[Zhang et al.(2018b)]{ZhangBB18b} Zhang B.-B., et al., 2018b, Nature Communications, 9, 447

\bibitem[Zhang et al.(2021)]{ZhangBB2021} Zhang B.-B., et al., 2021, Nature Astronomy, 5, 911

\bibitem[Zhang,Woosley \& MacFadyen(2003)]{ZhaWoo2003} Zhang W., Woosley S.~E., MacFadyen A.~I., 2003, \apj, 586, 356
\end{thebibliography}
\end{document}